\documentclass[fleqn,usenatbib]{mnras}

\usepackage{newtxtext,newtxmath}

\usepackage[T1]{fontenc}

\DeclareRobustCommand{\VAN}[3]{#2}
\let\VANthebibliography\thebibliography
\def\thebibliography{\DeclareRobustCommand{\VAN}[3]{##3}\VANthebibliography}

\usepackage{graphicx}	
\usepackage{amsmath}	

\usepackage{amsmath}
\usepackage{bm}
\usepackage{booktabs}
\usepackage{hyperref}
\usepackage{enumitem}

\newcommand{\dd}{\,\mathrm{d}}

\newcommand{\Hess}{\operatorname{Hess}}
\newcommand{\diag}{\operatorname{diag}}

\newcommand{\xixi}{\bm{\xi}}
\newcommand{\xx}{\bm{x}}
\newcommand{\Amat}{\mathsf{A}}
\newcommand{\gmat}{\mathsf{g}}

\newcommand{\vphi}{\varphi}
\newcommand{\xv}{\mathbf{x}}
\newcommand{\kv}{\mathbf{k}}
\newcommand{\gv}{\mathbf{g}}
\newcommand{\Of}{\mathcal{O}}
\newcommand{\norm}[1]{\left\lVert #1 \right\rVert}

\title[Fast(er)PM and Moving Mesh]{Fast(er)PM and Moving Mesh: JAX-native Geometric Multigrid Methods}

\author[B. Horowitz ]
{
Benjamin Horowitz$^{1,2,3}$\thanks{E-mail: ben.horowitz@ipmu.jp}
\\
$^{1}$Kavli IPMU (WPI), UTIAS, The University of Tokyo, Kashiwa, Chiba 277-8583, Japan \\
$^{2}$Center for Data-Driven Discovery, Kavli IPMU (WPI), UTIAS, The University of Tokyo, Kashiwa, Chiba 277-8583, Japan\\
$^{3}$Lawrence Berkeley National Lab, 1 Cyclotron Road, Berkeley, CA 94720, USA
}

\date{Accepted XXX. Received YYY; in original form ZZZ}

\pubyear{\the\year{}}

\begin{document}
\label{firstpage}
\pagerange{\pageref{firstpage}--\pageref{lastpage}}
\maketitle

\begin{abstract}

Efficient, differentiable Poisson solvers are a key component of modern particle--mesh simulations and field-level inference pipelines. FFT-based solvers are extremely effective on fixed Cartesian meshes, but they impose global all-to-all communication and rely on symmetries that are lost in adaptive or non-Cartesian coordinates. In this work, we present a JAX-native geometric multigrid framework for particle--mesh gravity and argue that multigrid plays two complementary roles: on fixed meshes it can be a competitive, communication-avoiding alternative to FFTs, while on moving meshes it becomes the enabling solver. For static FastPM evolution, warm-started Chebyshev multigrid acts as a defect-correction method, exploiting temporal coherence between time steps to reduce the number of V-cycles required for field-level accuracy. At large mesh sizes this reduces memory pressure and yields comparable or faster wall-clock performance than distributed FFTs, with up to a factor of two reduction in total GPU time at fixed final mesh size. We then embed the same solver in a differentiable moving-mesh particle--mesh method, where adaptive coordinate deformation produces a variable-coefficient curvilinear Poisson equation that cannot be solved by ordinary FFT diagonalization. The resulting method concentrates force resolution in nonlinear structures while retaining a regular, JAX-compilable, automatically differentiable array workflow. These results suggest geometric multigrid can be a practical bridge between fast fixed-grid PM methods and differentiable adaptive-force cosmological simulations.

\end{abstract}




\section{Introduction}

The Poisson equation \citep{poisson1826memoire} is the backbone of computational physics, providing the canonical route from a source field to a potential in problems ranging from gravity and electrostatics to pressure projection in fluid dynamics. In simulations, solving the Poisson equation often dominates the computational costs \citep{ibeid2020fft}, particularly when it must be performed repeatedly, at high resolution, or as part of an optimization/inference loop \citep{2021A&C....3700505M}. As scientific computing increasingly moves toward accelerator-based and differentiable programming frameworks, there is a growing need for Poisson solvers that are not only fast, but also flexible, composable, and compatible with automatic differentiation.

In the context of approximate $N$-body simulations, the current standard is to solve the Poisson equation on a grid using fast Fourier transforms. This approach was first introduced in \cite{eastwood1974shaping} and popularized in the astrophysics community in \cite{1981csup.book.....H} and \cite{1985ApJS...57..241E}. As the computational demands of cosmological dark matter simulations expanded, PM methods and their derivatives (TreePM, P3M, etc.) continued to rely on FFT methods to provide strong scaling in multi-node and distributed environments \citep{1997astro.ph.12217K,2005NewA...10..393M,fastPM}. 

More recently, this approach has become key in field-level inference pipelines. These approaches explore a high-dimensional parameter space of the simulation initial conditions (and optionally physical parameters) using gradient-based sampling or optimization methods. It is straightforward to calculate gradients of the various relevant operators either explicitly (e.g. \citet{2017JCAP...12..009S}) or via auto-differentiation libraries (e.g. \citet{2021A&C....3700505M}).
This method has been applied to FastPM \citep{fastPM} and its modern GPU derivatives \citep{2021A&C....3700505M,2024ApJS..270...36L} as well as COLA framework \citep{2013JCAP...06..036T} and its variants \citep{2024MNRAS.535.1258D,2025MNRAS.540..716M}. The gradient information has been used to reconstruct the initial condition in the local universe as traced by galaxies \citep{2018BORG,2025MNRAS.540..716M,2026arXiv260426823D} as well as the $z \sim 2$ universe traced by Lyman-$\alpha$ forest \citep{2019TARDIS,2022ApJS..263...27H}. However, these forward modeling frameworks are limited to the scales resolvable by fixed grid methods and the availability of forward models for the observable tracers \citep{2025MNRAS.544..960H}. 

Instead of solving the Poisson equation using FFTs, the problem can be viewed as a large system of linear equations and solved via multigrid methods \citep{fedorenko1962relaxation,brandt1977multi}. While early cosmological $N$-body simulations frequently used these methods (i.e. \citet{1995ApJS..100..269P,1997ApJS..111...73K}), FFT methods have largely superseded them due to scalability and the availability of highly optimized FFT packages of high performance computing (HPC) systems (most famously Fastest Fourier Transform of the West (FFTW) \citep{1998assp....3...34F,2005IEEEP..93..216F}). However, astrophysical fluid simulations continue to frequently use multigrid methods due to their natural application to multiscale adaptive mesh refinement problems (i.e. \citet{2018JPhCS1031a2021R,2020ApJS..249....4S,2026arXiv260503563K}). 

While the current multi-node CPU architecture seems to clearly favor FFT methods in the static mesh case \citep{ibeid2020fft},  it is far from clear if these results translate to modern GPU architectures. As briefly explored in \citet{2025arXiv251213403H} (henceforth \textsc{DiffHydro}), the all-to-all communication required for distributed FFTs is not inherently "GPU friendly" and MG methods can have superior performance to the standard JAX FFT libraries. More recently, \textsc{jaxdecomp}  \citep{kabalan2026jaxdecomp} has provided wrappers for \textsc{cuDecomp} \citep{romero2022distributed} which have superior asymptotic performance compared to existing FFT libraries. This motivates re-examining the tradeoffs between FFT and multigrid methods for solving the Poisson equation in the context of cosmological simulations.

Beyond speed, the multigrid approach allows for Poisson solving in arbitrary coordinates. As shown in \cite{1995ApJS..100..269P}, continuous deformations of the mesh (i.e. quasi-Lagrangian) can enable adaptive force resolution with modest additional computational cost. Unlike methods like TreePM, this approach requires only vector/array operations and can be differentiated directly with JAX autodiff. This provides a natural bridge between fixed-grid PM solvers and adaptive force resolution, while retaining the same array-programming structure used in \textsc{diffhydro}.

In this work, we revisit the use of geometric multigrid methods for solving the Poisson equation in particle--mesh cosmological simulations. In Section~\ref{sec:methods}, we review the PM gravity solve and establish a common discrete formulation for the spectral and multigrid approaches. Section~\ref{sec:warmstart} introduces a warm-started, Chebyshev-smoothed multigrid solver for FastPM and compares its convergence, accuracy, memory requirements, and distributed-GPU performance with an FFT-based implementation. In Section~\ref{sec:moving}, we extend the same framework to continuously deformed coordinates, present a differentiable moving-mesh PM scheme, and test its accuracy in both forward simulations and field-level reconstruction. We discuss the resulting accuracy--performance trade-offs and summarize our conclusions in Section~\ref{sec:disc}. Additional tests of error accumulation, details of the Jacobi smoother and grid-transfer operators, and the moving-mesh limiters are presented in Appendices~\ref{app:accum}, and \ref{sec:limiters}.

\section{Methods}
\label{sec:methods}
\begin{figure*}
    \centering
    \includegraphics[width=0.99\linewidth]{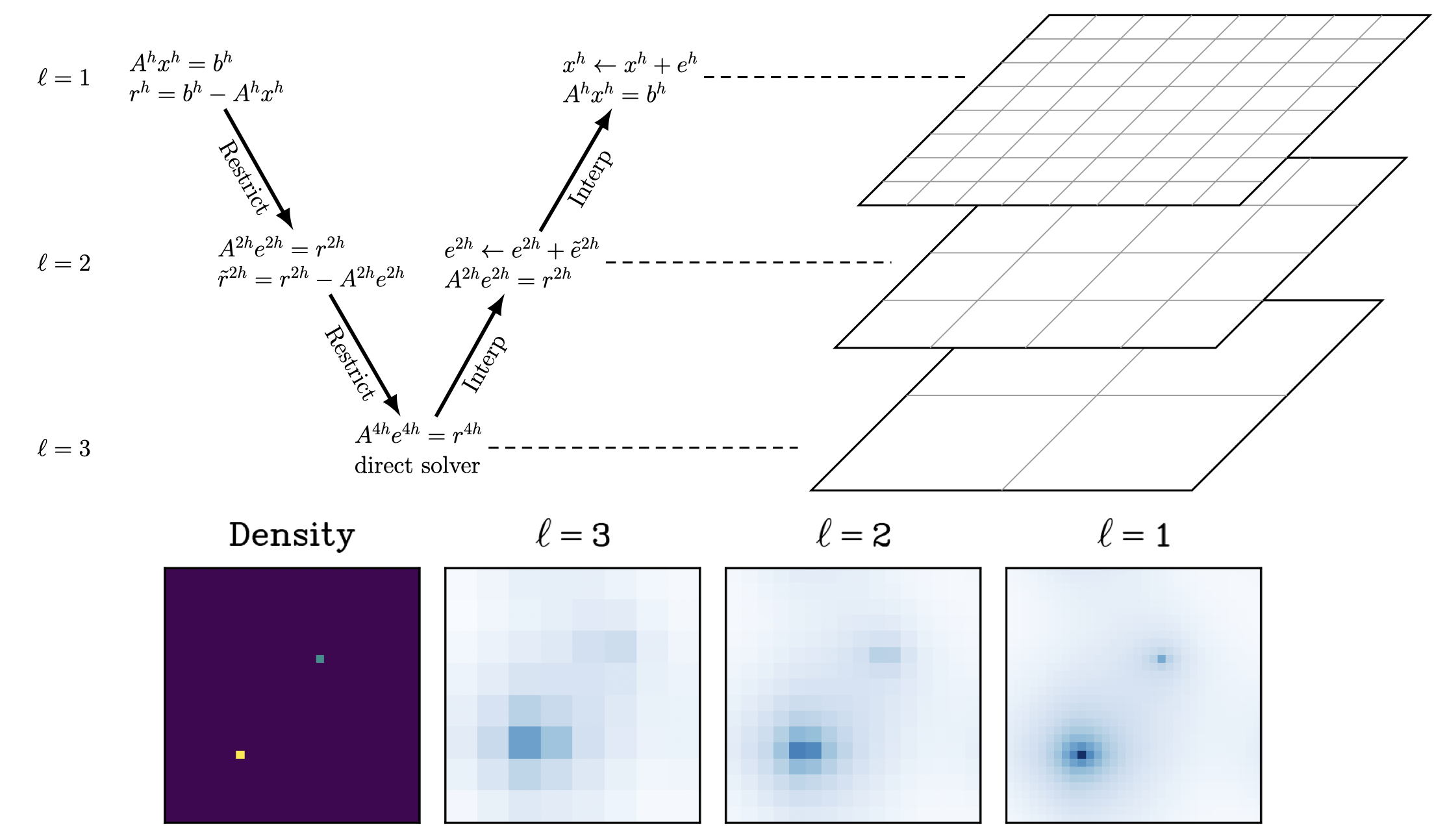}

    \caption{Geometric multigrid V-cycle for the Poisson solve. On the finest mesh, the potential satisfies $A^h\vphi^h=F^h$ and the residual is $r^h=F^h-A^h\vphi^h$. The residual is restricted to coarser meshes, where the error equations $A^{2h}e^{2h}=r^{2h}$ and $A^{4h}e^{4h}=r^{4h}$ are solved. The resulting corrections are prolongated back to the fine grid and added to the solution, $\vphi^h\leftarrow\vphi^h+e^h$. The lower panels show the same source represented across the mesh hierarchy, illustrating how coarse grids capture the long-wavelength component of the gravitational potential. From \citet{2025arXiv251213403H}.}
    \label{fig:mg_vcycle}
\end{figure*}

\subsection{The PM gravity solve}
On a periodic cubic mesh of side $N$ with spacing $h$, particles are deposited (e.g.\ by
cloud-in-cell interpolation) to form the density contrast $\delta(\xv)$. We write the source as
$F(\xv)\equiv\delta(\xv)$ and factors such as $4\pi G$, powers of the scale factor, or the
mean-density normalization can be absorbed into $F$ for the PM applications considered here. The
(comoving, suitably normalized) potential satisfies
\begin{equation}
  \nabla^2 \vphi = F,
  \label{eq:poisson}
\end{equation}
and the force per unit mass is $\gv = -\nabla\vphi$. Discretizing the Laplacian with the standard
second-order seven-point stencil yields the linear system
\begin{equation}
  A\,\vphi = F,
  \label{eq:linsys}
\end{equation}
where $A$ is symmetric negative-semidefinite with a one-dimensional null space spanned by the
constant mode (the discrete analogue of the requirement that a periodic source integrate to zero).
Solvability requires $\sum_\xv F(\xv) = 0$ which the FFT enforces implicitly by setting the
$\kv=0$ Fourier coefficient of the potential to zero, and in the multigrid solver we project the
mean out of $F$.

\subsection{Spectral (FFT) solution and its communication cost}
The FFT solves \eqref{eq:linsys} directly,
\begin{equation}
  \vphi = \mathcal{F}^{-1}\!\left[\hat F(\kv)\,/\,\lambda(\kv)\right],
  \label{eq:fftsolve}
\end{equation}
where $\lambda(\kv)$ are the eigenvalues of $A$ (for the discrete stencil,
$\lambda = -\tfrac{4}{h^2}\sum_i \sin^2(k_i h/2)$; the continuum choice $\lambda=-|\kv|^2$ is also
common). The cost is fixed with a forward transform of $F$, a pointwise multiply, and an inverse
transform. The force additionally requires the gradient; in the spectral representation this is a
pointwise multiply by $i\,\widehat{\partial_i}$ followed by an inverse transform per Cartesian
component, so a full force evaluation costs four transforms in total (one forward, three
inverse). On a distributed mesh each transform entails one or two global transposes, i.e.\
all-to-all communication, which dominates the wall-clock time at scale and is independent of any
prior knowledge of the solution.

\subsection{Geometric multigrid}
\label{sec:geommg}

We summarize the multigrid procedure qualitatively in Fig \ref{fig:mg_vcycle}. Following the notation of the \textsc{diffhydro} self-gravity solver, the flat-mesh PM Poisson
problem is written as
\begin{equation}
  A^h \vphi^h = F^h,
\end{equation}
where $h$ is the mesh spacing on the finest level, $A^h$ is the second-order seven-point discrete
Laplacian, $F^h$ is the mean-subtracted source, and $\vphi^h$ is fixed to have zero spatial mean.
Multigrid does not attempt to diagonalise $A^h$. Instead, it solves this linear system by repeatedly
correcting an approximate solution using information from a hierarchy of coarser meshes with
spacings $2h,4h,\ldots$. The key observation is that local relaxation methods remove oscillatory
error components quickly, while smooth errors are slow to decay on the fine grid but become
oscillatory after restriction to a coarser grid.

For a current iterate $\vphi^{h,(m)}$, the production smoother used in this work is a
Chebyshev-accelerated Jacobi smoother, motivated by discussions in literature (particularly \citet{2003JCoPh.188..593A,2011SJSC...33.2864B}). Chebyshev-accelerated smoothers differ from the classic damped-Jacobi (i.e. as done in \textsc{DiffHydro}) by specifically targetting the highest frequency modes in a controlled way. We apply a low-degree polynomial in the Jacobi-preconditioned operator,
\begin{equation}
  \vphi^{h,(m+v)} = \vphi^{h,(m)}
  + p_v\!\left((D^h)^{-1}A^h\right)
  (D^h)^{-1}\!\left(F^h-A^h\vphi^{h,(m)}\right),
  \label{eq:flat_chebyshev}
\end{equation}
where $D^h=\diag(A^h)$ and $p_v$ is a degree-$v$ Chebyshev polynomial chosen to damp the
high-frequency part of the error spectrum. In practice, we target the upper part of the spectrum of
the Jacobi-preconditioned periodic Laplacian, using the interval $[2/\alpha,2]$ with
$\alpha\simeq 8$. We note that the ``optimal" choice of $\alpha$ is problem specific. This gives a stronger smoother at nearly the same communication cost as $v$
Jacobi sweeps, because the entire polynomial can be evaluated inside the same wide-halo region.

Weighted Jacobi is retained as the simpler secondary smoother and as a useful ablation. In that
case one uses
\begin{equation}
  \vphi^{h,(m+1)} = \vphi^{h,(m)}
  + \omega\,(D^h)^{-1}\!\left(F^h-A^h\vphi^{h,(m)}\right),
  \label{eq:flat_jacobi}
\end{equation}
with $\omega=2/3$ for the three-dimensional periodic Laplacian. This Jacobi version has the same
local stencil structure and fits naturally into the same multigrid hierarchy, but it damps the
oscillatory error less efficiently per level visit. We therefore use Chebyshev smoothing as the
default in all production timings and quote Jacobi primarily as the baseline solver.

The smoother, whether Chebyshev or Jacobi, is applied for a small number of
pre-smoothing steps to damp cell-scale error and produce the fine-grid residual
\begin{equation}
  r^h = F^h - A^h\vphi^h .
  \label{eq:flat_residual}
\end{equation}
The residual is then restricted to the next coarser mesh,
\begin{equation}
  r^{2h} = R r^h,
\end{equation}
and the coarse-grid problem is posed as an error equation,
\begin{equation}
  A^{2h} e^{2h} = r^{2h} .
  \label{eq:coarse_error}
\end{equation}
After solving this equation approximately, recursively or directly on the coarsest level, the coarse
error is prolongated and added back to the fine solution,
\begin{equation}
  \vphi^h \leftarrow \vphi^h + P e^{2h},
  \label{eq:coarse_correction}
\end{equation}
followed by several post-smoothing sweeps to remove interpolation-scale error introduced by the
correction. Here $R$ is full-weighting restriction and $P$ is trilinear prolongation. In the static
uniform-mesh case we rediscretise the same seven-point operator on each level. In the moving-mesh
case (see Sec.~\ref{sec:moving}), the geometry coefficients are restricted to each level before constructing the
corresponding coarse operator.


For the model Poisson problem the V-cycle reduces the error by a mesh-independent factor
$\rho\in(0,1)$ per cycle,
\begin{equation}
  \norm{\vphi^{(k)} - \vphi^\star} \;\le\; \rho^{\,k}\,
  \norm{\vphi^{(0)} - \vphi^\star},
  \label{eq:conv}
\end{equation}
with $\rho$ typically of order $0.1$ for a well-tuned cycle. The work per level decreases by a factor
of eight in three dimensions, so the total number of cell operations in a complete hierarchy is
$N^3(1+1/8+1/8^2+\cdots)\simeq(8/7)N^3$, yielding the optimal $\mathcal{O}(N^3)$ complexity. This
is the central algorithmic distinction from the FFT solver; the FFT has a fixed
$\mathcal{O}(N^3\log N)$ work profile and a global communication pattern, whereas multigrid has a
local stencil work profile and a tunable residual tolerance.

On a distributed (i.e. ``sharded") mesh, the smoother and transfer stencils require only the values in a thin boundary
layer of each subdomain. These are supplied by halo exchanges: point-to-point messages between
geometrically neighboring processes whose volume is proportional to the surface area of the local
subdomain. This nearest-neighbor pattern is the structural advantage of multigrid over the FFT's
all-to-all transposes on large process grids. The coarsest levels, where a subdomain holds only a few
cells, are gathered to a replicated coarse grid and solved redundantly without further inter-node
communication (see \S\ref{sec:impl}).

Equation~\eqref{eq:conv} also explains why multigrid is especially attractive for time-stepped PM
evolution. The number of cycles required to reach a target tolerance depends on the quality of the
initial guess $\vphi^{(0)}$. A cold solve starts from $\vphi^{(0)}=0$, but a dynamical PM calculation
provides a nearly free predictor from the previous time step. The rest of this work exploits this
``warm-start'' structure wherein multigrid is used as a defect-correction method for the small residual left
by an already good potential estimate.

\section{Fast(er)PM with Warm-start Chebyshev Multigrid}
\label{sec:warmstart}

In this section, we compare an optimized multigrid solver to the FFT approach \citep{kabalan2026jaxdecomp}.
The PM evolution below uses the multigrid hierarchy as a warm-started defect-correction solver, with Chebyshev smoothing as the default level smoother. The role of the smoother is important since warm starting reduces the amplitude of the residual, while Chebyshev smoothing removes the remaining high-frequency error more effectively than the damped-Jacobi baseline (as done in DiffHydro) without changing the local communication pattern of the V-cycle. 

\subsection{Temporal coherence and the initial guess}
An iterative solver returns
$\vphi^\star$ to a tolerance after a number of steps that decreases as $\vphi^{(0)}$ improves. In a
time-stepped PM simulation, the potential at step $n$ differs only slightly from that at step
$n\!-\!1$, because the density evolves continuously and the time step is bounded by accuracy and
stability of the integrator. The previous potential is therefore an excellent initial guess. If the
relative inter-step change is $\epsilon = \norm{\vphi_n-\vphi_{n-1}}/\norm{\vphi_n}$, then from the
cold start $\vphi^{(0)}=0$ the initial relative error is unity and reaching tolerance $\tau$ needs
$k_{\rm cold}\approx \log\tau/\log\rho$ cycles, whereas from the recycled start the initial error is
$\Of(\epsilon)$ and
\begin{equation}
  k_{\rm warm} \;\approx\; \frac{\log(\tau/\epsilon)}{\log\rho}
  \;=\; k_{\rm cold} - \frac{\log(1/\epsilon)}{\log(1/\rho)} .
  \label{eq:kwarm}
\end{equation}
For $\epsilon$ of a few percent and $\rho\sim0.1$ this removes one to two V-cycles at fixed
tolerance. In practice a single warm Chebyshev-smoothed V-cycle reaches the accuracy that a cold
solve needs several cycles to reach (\S\ref{sec:results}), so the recycled start together with the
stronger Chebyshev smoother is the operating point that makes multigrid competitive with the FFT on
a per-step basis.

\begin{figure*}
    \centering
    \includegraphics[width=0.99\linewidth]{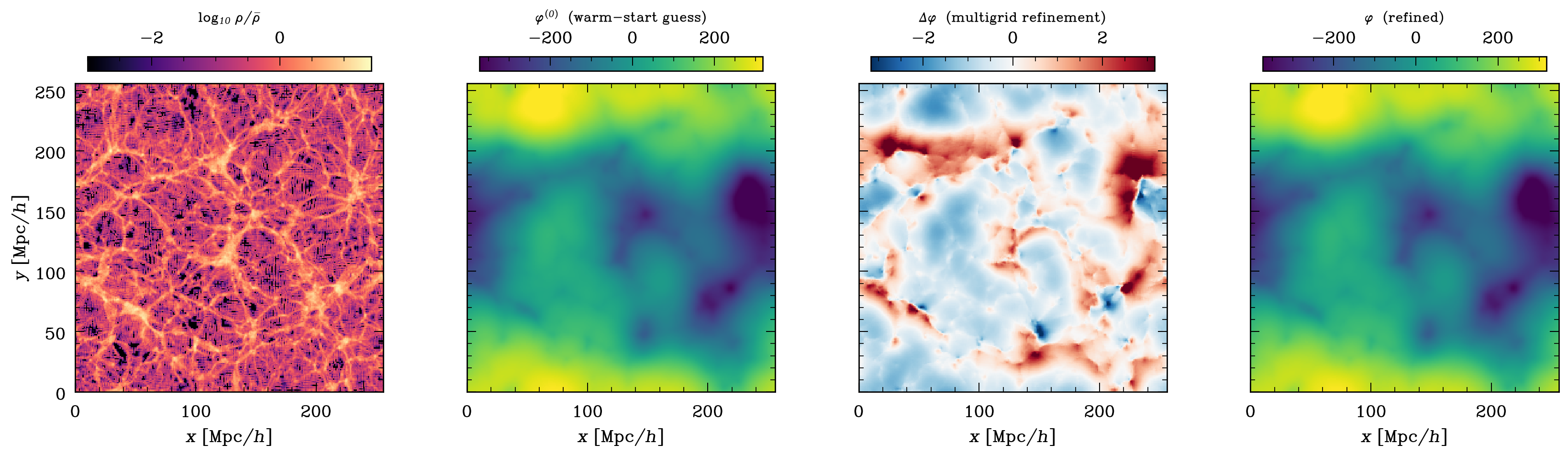}

    \caption{Warm-start Chebyshev multigrid as defect correction in a $256^3$ particle--mesh solve. From left to right: the logarithmic density contrast, the initial potential guess $\vphi^{(0)}$ recycled from the previous step, the multigrid correction $\Delta\vphi$, and the refined potential $\vphi=\vphi^{(0)}+\Delta\vphi$. The correction is small compared with the full potential and is concentrated around nonlinear structure, showing why a warm-started Chebyshev-smoothed cycle can converge rapidly.}
    \label{fig:warm_refinement}
\end{figure*}

\begin{figure}
    \centering
    \includegraphics[width=0.99\linewidth]{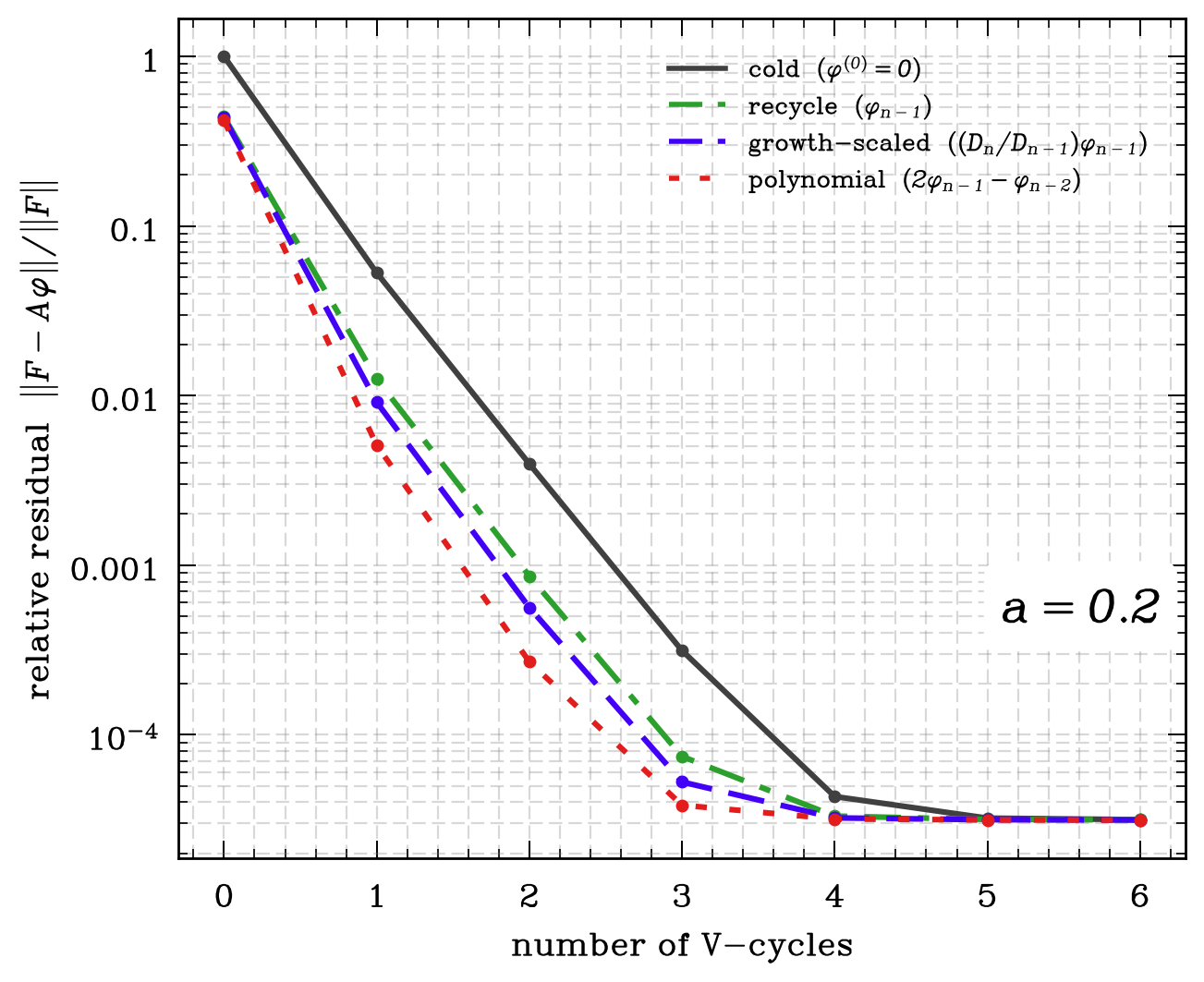}
    \centering
    \includegraphics[width=0.99\linewidth]{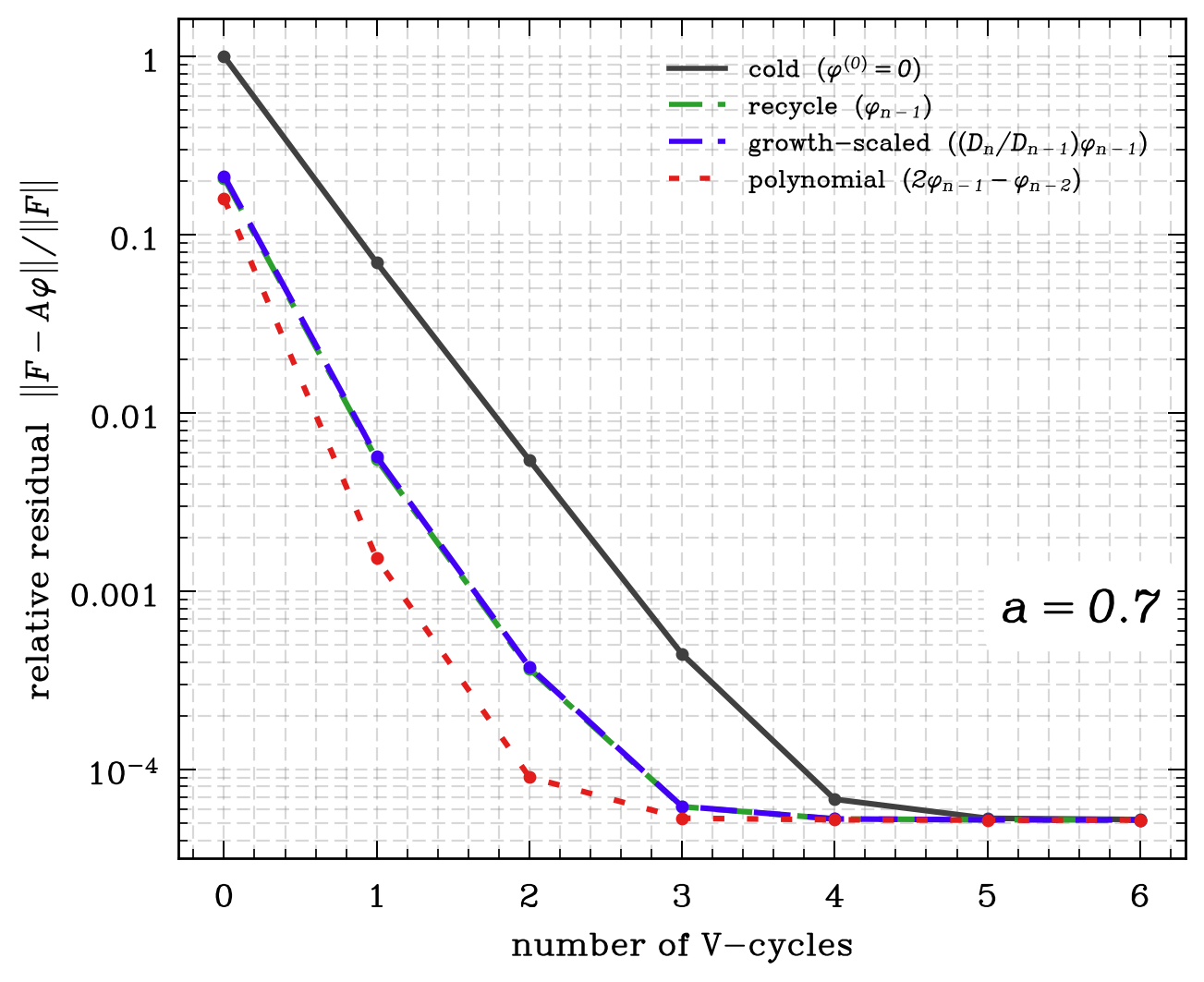}
    \caption{Residual convergence for cold and warm Chebyshev-multigrid solves at early ($a=0.2$) and late ($a=0.7$) times. Curves show the relative residual $\norm{F-A\vphi}/\norm{F}$ after each V-cycle for a cold start, simple recycling, growth-scaled recycling, and polynomial extrapolation. Warm-start predictors reduce the initial residual by orders of magnitude and typically remove one to two V-cycles at fixed tolerance.}
    \label{fig:warm_convergence}
\end{figure}
\subsection{Predictors}
We consider three initial guesses of increasing sophistication.\newline

\textbf{(i) Recycling.} $\vphi^{(0)}_n = \vphi_{n-1}$. Simple, requires no extra storage beyond
the previous potential, and already captures the bulk of the solution.

\textbf{(ii) Growth-scaled predictor.} The Poisson operator \eqref{eq:linsys} is linear, so a
uniform rescaling of the source rescales the solution identically. In the linear regime the density
contrast grows as the linear growth factor $D(a)$, $\delta \to (D_n/D_{n-1})\,\delta$, and hence
\begin{equation}
  \vphi^{(0)}_n = \frac{D(a_n)}{D(a_{n-1})}\,\vphi_{n-1}
  \label{eq:growth}
\end{equation}
exactly removes the linear-growth component of the inter-step change, leaving only the
nonlinear, mode-coupling residual for the cycles to resolve. This predictor is specific to
cosmological growth and costs only a scalar multiply (i.e. can be done in-place).

\textbf{(iii) Polynomial extrapolation.} With more history, a linear (or quadratic)
extrapolation $\vphi^{(0)}_n = 2\vphi_{n-1}-\vphi_{n-2}$ ($\,3\vphi_{n-1}-3\vphi_{n-2}+\vphi_{n-3}$)
captures steady drift of the potential and can further reduce $\epsilon$ when the trajectory is
smooth (see also \citet{2018JPhCS1031a2021R}).\newline

Warm-starting is mathematically a defect correction. Writing $\vphi=\vphi^{(0)}+e$, the error
satisfies $A\,e = \delta - A\vphi^{(0)} \equiv r^{(0)}$, the residual of the initial guess. Because
$\vphi^{(0)}$ is accurate, $r^{(0)}$ is small, and the multigrid cycles converge it rapidly, recovering $\vphi=\vphi^{(0)}+e$. This framing makes explicit that the solver operates on a small
residual quantity which is numerically favorable. When a guaranteed monotone decrease of the residual
is desired (e.g.\ for tight tolerances or stiff configurations), the same $\vphi^{(0)}$ can seed a
multigrid-preconditioned conjugate-gradient iteration, with multigrid as the preconditioner and the
recycled potential as the initial iterate. We show the fractional error as a function of cycle number for different predictors vs. cold start in Fig \ref{fig:warm_convergence}.

A potential concern with recycling is the accumulation of solver error along the trajectory. Each
step's force is computed from an approximate potential, perturbing the particle positions and hence
the next step's source. As explored further in Appendix \ref{app:accum}, the integrator's own truncation
error sets the accuracy floor, provided each warm solve reduces the residual below this floor (a
small, fixed number of cycles), the per-step solver error does not accumulate coherently, and the
final field tracks the reference (FFT) solution to a controlled tolerance set by the chosen cycle
count (\S\ref{sec:results}). The first step has no predecessor and is initialized once with an
accurate solve, in our case a deep multigrid cycle used only at initialization.

\subsection{Cost, communication, and memory}
\label{sec:cost}

Per force evaluation, the spectral path performs four global transforms (one forward, three inverse
for the gradient components), i.e.\ of order eight global transposes. The multigrid path performs
one warm Chebyshev-smoothed solve consisting of one or two V-cycles (each dominated by halo exchanges
proportional to subdomain surface area) followed by a real-space finite-difference gradient that
reuses the same local halo machinery and requires \emph{no} global communication. The communication advantage of
multigrid is therefore twofold. It replaces all-to-all transposes by nearest-neighbor exchanges,
and it obtains the gradient locally rather than by three additional inverse transforms. In terms of
transform-equivalent operation count the second factor alone is a fixed $\sim\!4\times$ reduction
(four transforms replaced by a single warm solve plus a local stencil) that applies even at modest
scale. The first factor (all-to-all versus halo) grows with node count. We stress, however, that
these are operation-count and communication-pattern arguments, not the measured wall-clock ratio.
The realized per-step speedup is a more modest $\sim\!1.3\text{--}1.6\times$ in the regime where
multigrid wins (\S\ref{sec:results}), because the FFT is a highly optimized library kernel and its
all-to-all, while pattern-unfavorable, is fast on a well-provisioned fabric/interconnect. The wall-clock gap is
also sensitive to run-to-run variance in the global collective, so we quote it only from matched
single-allocation measurements.

The spectral solve operates in complex arithmetic and materializes transposed copies of the field
during the transform, so its transient footprint is several times the real mesh; in particular the
distributed transform allocates pencil-transpose scratch of order a full $[N^3,3]$ vector field per
axis. The Chebyshev multigrid solve operates entirely in single-precision real arithmetic on the
mesh hierarchy (whose total size is $\tfrac{8}{7}N^3$) and communicates only thin halo frames, so its
transient footprint is substantially smaller. The practical consequence is a lower memory floor so that,
after the initial conditions are generated once (a single deep solve/FFT), the warm-start
Chebyshev-multigrid evolution requires \emph{no} FFT and can be run on roughly half the node count
that the spectral path needs to hold the same mesh. On the $40\,$GB A100 partition of Perlmutter we find the
FFT path first fits a $2048^3$ mesh at $64$ GPUs ($16$ nodes), whereas the FFT-free multigrid
evolution fits at $32$ GPUs ($8$ nodes), and a $1024^3$ mesh that the FFT needs two nodes for runs
on a single node with multigrid (\S\ref{sec:results}). Because the memory floor is set by the
solver-independent particle state (two full $[N^3,3]$ position/momentum fields), reaching still
lower node counts is a state-storage problem rather than a solver choice.

At fixed mesh size, increasing the node count shrinks each subdomain. The FFT all-to-all involves
more participants and smaller messages (latency-bound, scaling unfavorably), whereas Chebyshev
multigrid's halo exchanges remain nearest-neighbor and its per-node arithmetic falls. The exact
performance will depend strongly on the interconnect between GPUs, with NVIDIA Infiniband providing
significant scaling advantage for FFT tasks \citep{kabalan2026jaxdecomp}. The warm-start Chebyshev
advantage is therefore expected to grow with interconnects for which all-to-all
collectives are relatively more expensive than nearest-neighbor halo exchanges.

\begin{figure}
    \centering
    \includegraphics[width=0.99\linewidth]{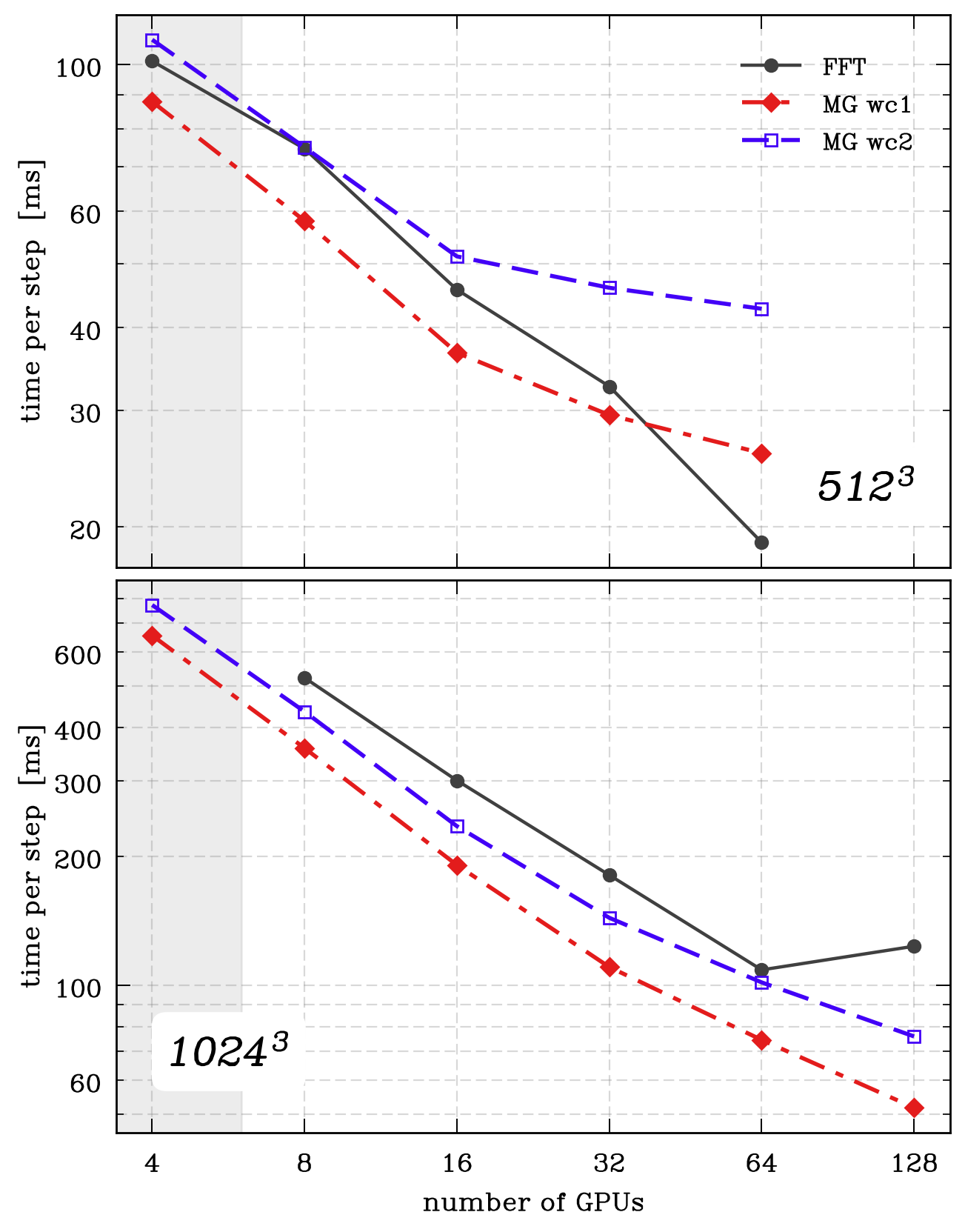}

    \caption{Strong-scaling comparison of FFT and warm-start Chebyshev-multigrid PM steps, measured in a single
    allocation on Perlmutter A100 ($40\,$GB) nodes. The panels show wall-clock time per step for
    $512^3$ and $1024^3$ meshes as the number of GPUs is varied. The FFT path is dominated by
    distributed transforms and global transposes, whereas warm Chebyshev multigrid uses one (wc1) or two (wc2) V-cycles plus local finite-difference forces. For the $1024^3$ mesh the one-cycle warm solve is faster
    than the FFT across the whole range, by $1.5\text{--}1.6\times$ in the sweet spot ($8$--$32$
    GPUs) and up to $2.4\times$ at $128$ GPUs where the FFT all-to-all becomes latency-bound; the
    multigrid point at $4$ GPUs has no FFT counterpart because the transform does not fit on a single
    node. For the smaller $512^3$ mesh the advantage is present at low GPU count but inverts by
    $64$ GPUs, where each multigrid subdomain reaches a halo-latency floor while the FFT keeps
    scaling.}
    \label{fig:mg_scaling}
\end{figure}

\begin{figure}
    \centering
    \includegraphics[width=0.99\linewidth]{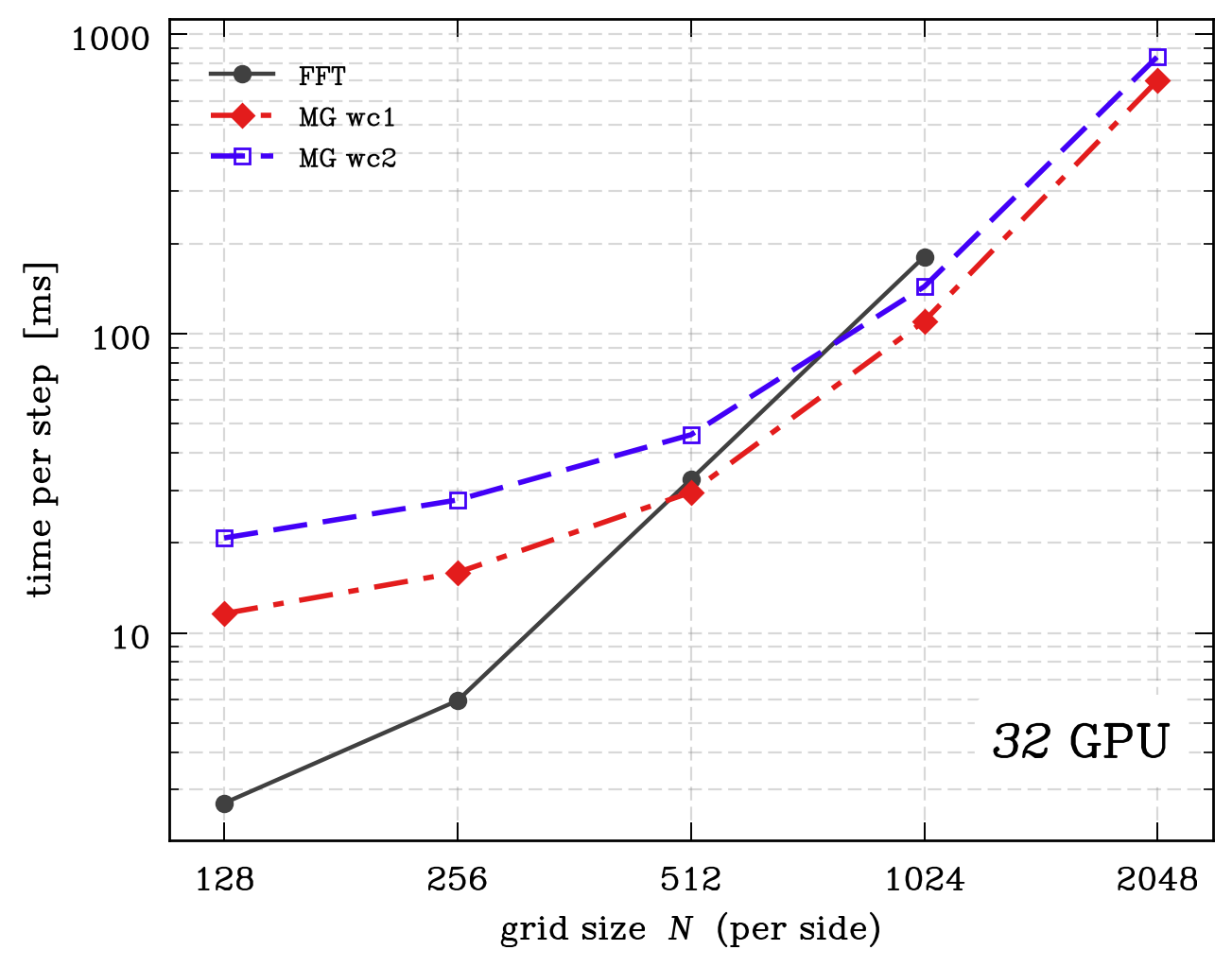}
    \caption{Per-step wall-clock time as a function of mesh size at fixed node count ($32$ GPUs),
    isolating the effect of the per-GPU problem size on the FFT/multigrid crossover. At small meshes
    the FFT is far cheaper (the warm solve is dominated by fixed halo-latency overhead), but the
    ratio climbs monotonically with mesh size crossing unity just above $512^3$. The
    $2048^3$ point has no FFT point because the spectral transform exhausts device memory at $32$ GPUs
    while the multigrid step still fits.}
    \label{fig:mg_gridscan}
\end{figure}

\begin{figure}
    \centering
    \includegraphics[width=0.99\linewidth]{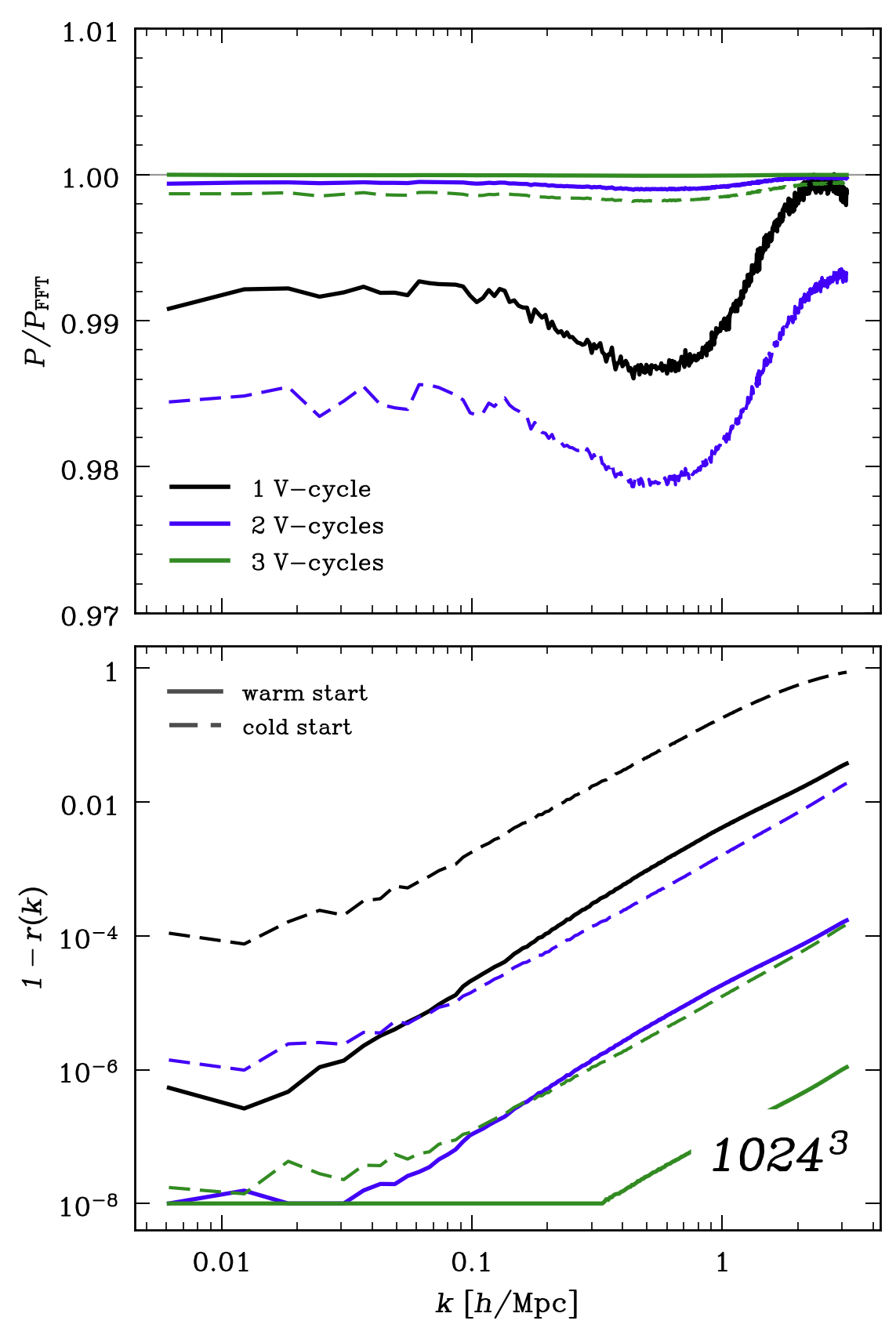}

    \caption{Field-level accuracy of the warm-started Chebyshev-multigrid PM evolution relative to an FFT reference for a $1024^3$ run. The top panel shows the power-spectrum transfer function $P/P_{\rm FFT}$, and the bottom panel shows the stochasticity $1-r(k)$. Solid curves use warm-started solves, while dashed curves show cold-start solves with the same number of V-cycles. Additional cycles systematically reduce both transfer-function bias and stochasticity, with the remaining error concentrated at high wavenumber and in collapsed regions. Note, the cold start, 1 V-cycle result has an average $P/P_{\rm FFT}$ of $\sim0.75$ and is below the $y$-limit cutoff.}
    \label{fig:pk_accuracy}
\end{figure}

\subsection{Implementation}
\label{sec:impl}
We implement both solvers in JAX with distributed transforms and halo exchanges provided
by \textsc{jaxdecomp} \citep{kabalan2026jaxdecomp}, within the \textsc{JaxPM} particle--mesh
framework. The mesh and particles are sharded over a two-dimensional device grid (a ``pencil'' decomposition) with the
third axis kept local (following the current implementation \textsc{jaxdecomp} for halo-exchange). Decomposing across all three dimensions is possible (i.e. ``cubic'' decomposition), but increased the surfaces to exchange from 4 to 6 per sub-volume and was found to be suboptimal for both FFT \citep{romero2022distributed} and our MG method.

The whole V-cycle is staged into a single compiled executable, and every
inter-device data movement is expressed as an explicit halo exchange over that pencil. Several implementation choices that matter for end performance are listed below. The first three are
the additions that give the present solver its speed over the originally published version in \textsc{DiffHydro}, and all
are exposed as flags with the original Jacobi path preserved as a fallback.\footnote{Several originally attractive optimizations were implemented, benchmarked and rejected, and we note some here because
each fails for an instructive reason and completeness. Restriction via a single zero-padded transform
(\texttt{concatenate}) defeats the compiler's operator fusion and inflates the transient footprint
sixfold ($7.5\to44.8\,$GiB/device at $1024^3$/$4$-GPU), so we retain the per-sample roll pattern.
Truncating the warm V-cycle to a few coarse levels (on the assumption that the recycled residual is
purely high-frequency) degrades the correlation with the reference ($r(k_{\rm Nyq}/2)$ collapses
from $0.76$ to $0.32$ at $512^3$), showing that the warm residual retains a genuine long-wavelength
component that the full hierarchy must resolve. Sending halo messages in \texttt{bfloat16} introduces a $\sim\!3.5\times10^{-4}$ residual noise floor that sits inside the convergence range, so exchanges are kept in single precision.}

\begin{itemize}\itemsep2pt
  \item \textbf{Communication-avoiding, whole-level fusion.} The smoother, residual, restriction, and prolongation operations on each level all need access to overlapping regions of the sharded field. Rather than exchange halos separately for each operation, we use a single wider exchange and reuse the data across the level. In the production implementation, this also allows us to apply a degree-$v$ Chebyshev smoother within the same fused level visit. A halo of width $v_1+2$ is sufficient to combine the $v_1$ pre-smoothing operations with the fused 27-tap full-weighting restriction, using just one exchange of the iterate and one of the source. We use a similar strategy on the way back up the hierarchy, where prolongation and post-smoothing share a single coarse-correction exchange. In total, this reduces the number of halo exchanges per level visit from five to two on the fine leg, plus one on the coarse leg. Since each sharded exchange requires a full pad-and-copy pass over the array, this cuts not only communication latency but also high-bandwidth-memory traffic. For a $1024^3$ mesh on a single node, the isolated Poisson solve drops from $226$ to $102\,\mathrm{ms}$ for a warm one-cycle solve, and from $918$ to $391\,\mathrm{ms}$ for a cold four-cycle solve, corresponding to a $2.2\text{--}2.4\times$ speedup. The trade-off is a modest increase in temporary storage ($\sim\!3\text{--}5\,\mathrm{GiB/device}$), so we enable the fused implementation only when sufficient memory is available ($\gtrsim\!32$ GPUs) and retain the original roll-based implementation when memory is constrained. 
  \item \textbf{Chebyshev smoothing.} Chebyshev smoothing is the default used in the
         multigrid solver, as described in Sec. \ref{sec:geommg}
. We apply a degree-$v$ Chebyshev polynomial in the
        Jacobi-preconditioned $7$-point operator, targeting the upper part of the spectrum on the
        interval $[2/\alpha,2]$ with $\alpha\approx8$. This gives a strictly better smoother per
        cycle at only $\sim\!5\%$ additional cost relative to damped Jacobi, and it fits inside the
        same wide-halo chunk as the Jacobi sweeps, so it inherits the same communication-avoiding fusion
        above. At the high-accuracy operating point ($2048^3$, two warm cycles), Chebyshev smoothing
        lowers the field-level decorrelation from the FFT reference by roughly an order of
        magnitude: $1-r(k_{\rm Nyq}/2)$ falls from $1.4\times10^{-2}$ for the Jacobi baseline to
        $1.4\times10^{-3}$. Because the fused Chebyshev path is also $1.27\times$ faster than the
        original Jacobi implementation in \textsc{DiffHydro}, the production solver is simultaneously faster and more
        accurate than the original damped-Jacobi version.
        
          \item \textbf{One-exchange finite-difference gradient.} The gravitational force ($\gv=-\nabla\vphi$) is formed
        from a single width-$2$ halo exchange feeding a fourth-order stencil, replacing twelve
        sharded \texttt{roll} operations (each of which is itself a collective permute). This removes
        roughly eight collective rounds and about ten temporaries per force evaluation and is
        mathematically identical to the roll-based gradient. It is the local-gradient step that makes the
        multigrid force free of any global communication once the potential on device.
         \item \textbf{Coarse-grid agglomeration.} Once a coarse level is small enough that each subdomain
        holds only a few cells, the grid is gathered to a replicated copy and the remaining levels
        are solved redundantly on each device, eliminating latency-bound exchanges on the deepest
        levels and restoring the long-wavelength accuracy that heavy sharding would otherwise
        truncate.
      \item \textbf{Whole-program compilation.} Staging the entire V-cycle into a single compiled
        executable, rather than dispatching each kernel eagerly, is essential. In our tests it
        accounted for a $5\text{--}6\times$ reduction in multigrid run time. Timings throughout
        exclude the one-time compilation via a warm-up call.
 
\end{itemize}

\subsection{Results}
\label{sec:results}
We evaluate on NVIDIA A100 GPUs (40Gb)\footnote{Perlmutter
hosts both $40\,$GB and $80\,$GB A100 nodes, and an unpinned allocation may mix the two, which
corrupts both memory-floor and timing measurements (the $80\,$GB nodes also carry $\sim\!1.3\times$
the memory bandwidth). All timings reported here pin the node class explicitly and measure the FFT
and multigrid solvers within a single allocation, because the FFT's global all-to-all is
sensitive to run-to-run fabric contention and splicing points across allocations can inflate the
apparent speedup severalfold. We verified with NCCL transport logging that the inter-node path uses
GPUDirect RDMA over the Slingshot fabric as intended, so the residual timing variance is inherent
collective cost and shared-fabric contention rather than a misconfiguration.} on the Perlmutter computing cluster. Notably, Perlmutter uses the
HPE Cray Slingshot interconnect between GPU nodes, which is likely sub-optimal for all-to-all
communication (i.e. other HPC systems may have better performance for FFT methods). All comparisons use identical initial conditions for each solver, with the multigrid
solver configured to the same discrete operator as the spectral reference so that differences
reflect solver convergence rather than discretization. Accuracy is quoted as the relative
$L_2$ difference of the final density field against the FFT solution. We note again that we are using the \textsc{jaxdecomp}  \citep{kabalan2026jaxdecomp} wrappers for highly optimized \textsc{cuDecomp} \citep{romero2022distributed} for FFT, unlike the original \textsc{DiffHydro} that used the native \textsc{jax} FFT implementation.

Table~\ref{tab:sweep} reports a full $256^3$ PM run of 30 steps as the number of warm-start
Chebyshev-smoothed V-cycles is varied. Each warm cycle reduces the final-field error by roughly an
order of magnitude, and the warm scheme Pareto-dominates a cold cycle. At equal cost (three
cycles) warm-start reaches $3.7\times10^{-3}$ where a cold cycle reaches only $3.4\times10^{-2}$,
and two warm cycles match the cold three-cycle accuracy at lower cost. A single warm Chebyshev cycle
is the fastest configuration and can achieve $\leq 1 \%$ power spectra accuracy relative to spectral methods, but may still be too coarse for some applications (see Fig.~\ref{fig:pk_accuracy}).

\begin{table}
\centering
\begin{tabular}{lcc}
\toprule
configuration & cost (ms/step) & final-field rel.\ $L_2$ \\
\midrule
FFT (reference)        & 17.9 & --- \\
cold MG (3 cycles)     & 23.3 & $3.4\times10^{-2}$ \\
warm MG, 1 cycle       & 16.5 & $2.7\times10^{-1}$ \\
warm MG, 2 cycles      & 20.1 & $3.0\times10^{-2}$ \\
warm MG, 3 cycles      & 23.3 & $3.7\times10^{-3}$ \\
warm MG, 4 cycles      & 26.6 & $5.2\times10^{-4}$ \\
\bottomrule
\end{tabular}
\caption{Accuracy/speed trade-off versus warm Chebyshev cycle count, $256^3$ mesh, 30 steps,
single GPU, using the growth-scaled predictor \eqref{eq:growth}. Warm-start Chebyshev multigrid
dominates the cold cycle: same cost, roughly an order of magnitude better accuracy.}
\label{tab:sweep}
\end{table}

Table~\ref{tab:scale} reports the full force-evaluation cost per step for the $1024^3$ mesh as the
GPU count is varied, the regime in which warm Chebyshev multigrid is competitive. The one-cycle warm
Chebyshev solve is faster than the FFT at every node count where both fit, by $1.46\times$ at $8$
GPUs, rising to $1.64\times$ at $32$ GPUs, and reaching $2.38\times$ at $128$ GPUs where the FFT
all-to-all becomes latency-bound and its per-step time actually increases. At the same time
multigrid fits on a single node ($4$ GPUs), where the distributed transform does not fit at all. The
two-cycle warm Chebyshev solve, which is the configuration matched to the FFT reference at the field
level (see below), is roughly $1.2\text{--}1.3\times$ faster than the FFT over the $8$--$32$ GPU
range. This confirms that the crossover is governed by the
per-GPU problem size rather than the absolute node count: for the smaller $512^3$ mesh the one-cycle
advantage is present at low GPU count ($1.15$--$1.29\times$ up to $16$ GPUs) but inverts by $64$ GPUs
($0.73\times$), where each multigrid subdomain has shrunk to a halo-latency floor ($\sim\!26\,$ms)
while the FFT continues to scale down. Figure~\ref{fig:mg_gridscan} shows the same trend directly as
a mesh-size scan at fixed node count.

The comparison at $2048^3$ (see Table \ref{tab:mg_fft_performance}) is set by memory rather than speed. On the $40\,$GB partition the spectral
path first fits at $64$ GPUs ($706.5\,$ms/step), whereas the FFT-free multigrid evolution fits at
$32$ GPUs ($703\,$ms/step for one cycle, $844\,$ms/step for two). Measured in cumulative GPU time, the resource
to advance one step is then $45.2\,$GPU$\cdot$s for the FFT against $22.5$--$27.0\,$GPU$\cdot$s for
multigrid, a $1.7$--$2.0\times$ reduction. The same
pattern holds at $1024^3$, where multigrid on one node ($4$ GPUs) advances a step in $2.62\,$GPU$\cdot$s
against $4.18\,$GPU$\cdot$s for the FFT on the two nodes it minimally requires.

\begin{table}
\centering
\begin{tabular}{lcccc}
\toprule
GPUs & FFT & warm MG, 1 cyc & warm MG, 2 cyc & speedup \\
\midrule
4    & (OOM) & 655.8 & 772.7 & --- \\
8    & 522.5 & 357.9 & 435.5 & $\mathbf{1.46\times}$ \\
16   & 300.7 & 190.8 & 235.2 & $\mathbf{1.58\times}$ \\
32   & 180.8 & 110.3 & 143.7 & $\mathbf{1.64\times}$ \\
64   & 108.6 & 74.5  & 101.5 & $\mathbf{1.46\times}$ \\
128  & 123.3 & 51.7  & 75.9  & $\mathbf{2.38\times}$ \\
\bottomrule
\end{tabular}
\caption{Full PM force-evaluation cost per step (paint, solve, gradient, read) in ms/step for a
$1024^3$ mesh as a function of GPU count, on Perlmutter A100 ($40\,$GB) nodes. Warm multigrid uses
the growth-scaled predictor with the fused, Chebyshev-smoothed solver of \S\ref{sec:impl} at 1 or 2
V-cycles. The FFT uses the matched discrete operator. Speedup is the FFT/multigrid wall-clock
ratio at equal GPU count for the one-cycle Chebyshev solve. FFT and multigrid were measured in the
same allocation to avoid cross-run network variance. The $4$-GPU row has no FFT entry because the
distributed transform exhausts single-node memory.}
\label{tab:scale}
\end{table}

\begin{table}
  \centering
  \caption{Timing comparison of the fused multigrid solver with Chebyshev smoothing against FFT and the Jacobi-smoothed multigrid implementation for a $2048^3$ box with 30 time-steps.}
  \label{tab:mg_fft_performance}
  \begin{tabular}{lccc}
    \toprule
    Method & GPUs & ms/step & GPU-s/step \\
    \midrule
    FFT
      & 64 & 706.5 & 45.2 \\
    MG Chebyshev, 1 cycle
      & 32 & 702.7 & 22.5 \\
    MG Chebyshev, 2 cycles
      & 32 & 844.4 & 27.0 \\
    MG Jacobi, 1 cycle
      & 32 & 814.7 & 26.0 \\
      MG Jacobi, 2 cycles
      & 32 & 1073.9 & 34.4 \\
    \bottomrule
  \end{tabular}
\end{table}

At the field level the warm-started Chebyshev-multigrid and FFT runs are visually indistinguishable.
The residual is sub-percent in relative $L_2$ and is localized to the dense cores of collapsed
halos, where the discrete operator and the finite cycle budget differ most. Quantitatively, at
$2048^3$ the two-cycle warm Chebyshev solve reproduces the reference power spectrum to
$P/P_{\rm FFT}=1.001$ and decorrelates by only $1-r(k_{\rm Nyq}/2)=1.4\times10^{-3}$, an order of
magnitude tighter than the original damped-Jacobi solver ($1.4\times10^{-2}$ at the same cycle
count) and, as noted above, at lower wall-clock cost. A single warm cycle is faster still but reaches only
$r(k_{\rm Nyq}/2)\approx0.80$, so two cycles remain the production choice where high field-level fidelity
is required. The cosmic-web structure (i.e. filaments, walls, and voids) is reproduced faithfully. As discussed in \citet{kabalan2026jaxdecomp}, care is required in the mass-assignment
halo width. At fixed box size, increasing the mesh resolution increases particle displacements
measured in cells, so an under-sized halo produces shard-boundary artifacts in the deposited field.
These are a property of the parallel deposition, common to both FFT and MG solvers, and are mitigated by sizing the exchange halo to the displacement scale.

\begin{figure*}
    \centering
\includegraphics[width=0.99\linewidth]{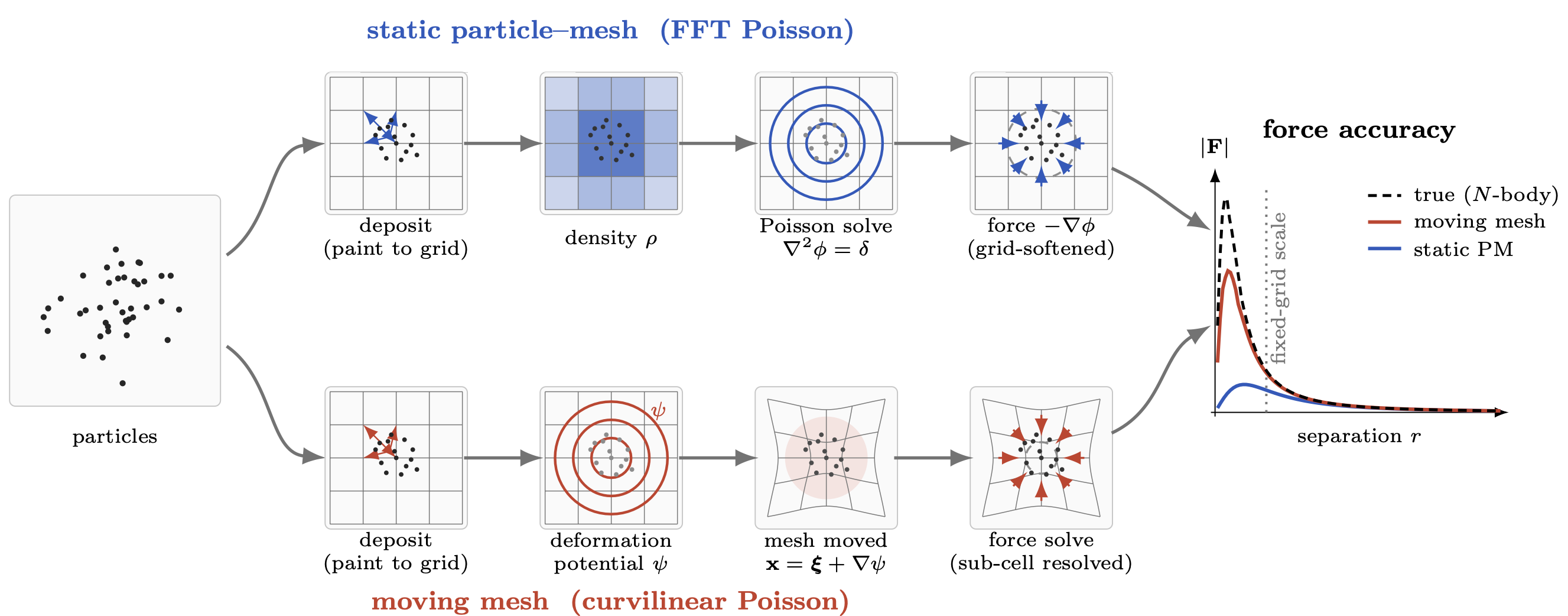}
    \caption{The standard static particle mesh compared with the moving mesh force calculation, starting from the same particle distribution
(left) and overlaid with the particles at every stage. In the static particle--mesh branch
(top) the particles are painted onto a fixed grid, the density $\rho$ is Fourier
transformed, $\nabla^2\phi=\delta$ is solved with an FFT, and the force $-\nabla\phi$ is
read back. The force resolution is set once by the grid cell (dashed circle). In the
moving-mesh branch (bottom) the same deposit drives a deformation potential $\psi$ whose
gradient contracts the mesh onto the particle overdensity,
$\mathbf{x}=\boldsymbol{\xi}+\nabla\psi$, so the subsequent curvilinear Poisson (requiring a multigrid method) solve
resolves the force on a much smaller cell (dashed circle) exactly where the particles
cluster. The right panel shows that the pairwise force
$|\mathbf{F}|(r)$ of the fixed grid is suppressed below its cell scale (blue), whereas the
moving mesh follows the true $N$-body force (dashed) to substantially smaller separations
(red). Both branches agree on large scales and differ only in how force resolution is
allocated.}
\label{fig:tikz_branches}
\end{figure*}
\section{Moving-Mesh Particle--Mesh Evolution}
\label{sec:moving}
\begin{figure*}
    \centering
    \includegraphics[width=0.99\linewidth]{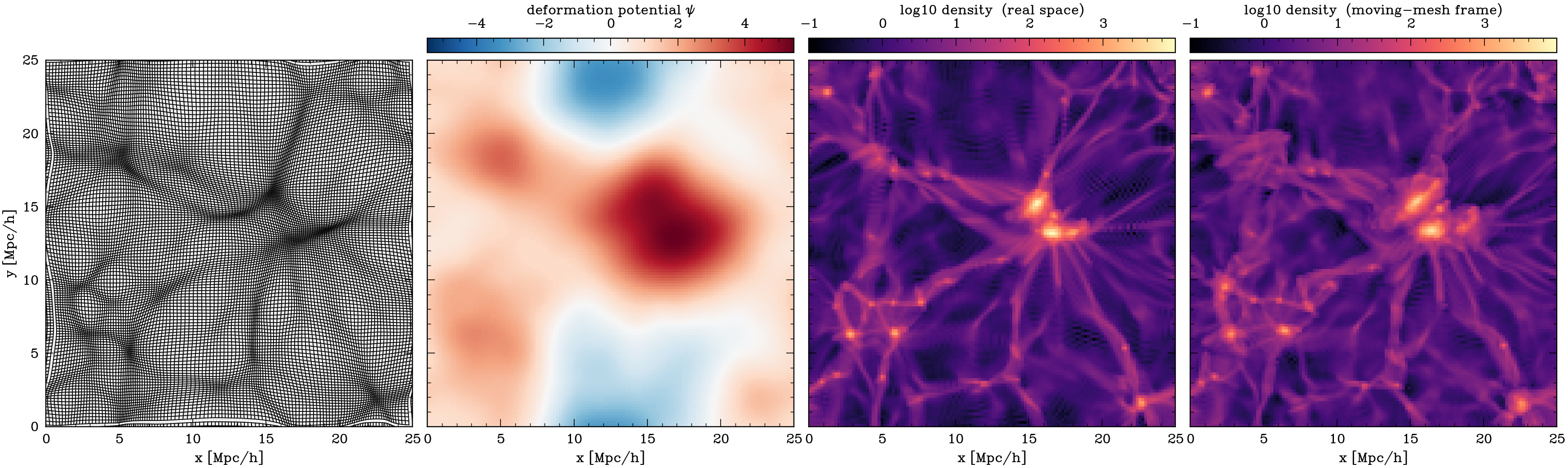}

    \caption{Example moving-mesh geometry in a cosmological field. Left: the deformed mesh in a slice through the volume, with cells contracted along filaments and around collapsed structures and expanded in voids. Middle left: the scalar deformation potential $\psi$ whose gradient defines the coordinate map $\xx(\xixi)=\xixi+\nabla\psi$. Middle right: the corresponding logarithmic density field in real space. Right: the logarithmic density field in the moving mesh frame. The figure illustrates how a single potential-flow deformation produces adaptive force resolution while preserving the Cartesian mesh topology. In addition, we see that the moving mesh frame works to partially linearize the field, more efficiently using the available grid elements for force calculation.}
    \label{fig:mesh_layer}
\end{figure*}

We next present a JAX implementation of the moving-mesh particle--mesh method based on earlier work in \citet{1995ApJS..100..269P,1998ApJS..115...19P}, summarized in Fig.~\ref{fig:tikz_branches} and shown qualitatively in Fig.~\ref{fig:mesh_layer}. The central idea is to compute the gravitational force on a mesh that deforms with the matter distribution which provides adaptive force resolution. Cells contract in overdense regions, increasing the local force resolution, and expand in voids, where high force resolution is less important.  As shown in the left two panels of Fig.~\ref{fig:mesh_layer}, the field has lower density contrast in the moving mesh frame. The method is therefore quasi-Lagrangian since the mesh approximately follows the particle distribution until shell crossing and nonlinear collapse, at which point explicit limiters prevent excessive distortion while preserving the underlying Cartesian topology. This method has also been used more recently for isobaric reconstruction \citep{2017MNRAS.469.1968P,2017ApJ...841L..29W} (alternatively known as nonlinear reconstruction \citep{2017ApJ...847..110Y,2018PhRvD..97d3502Z,2019ApJ...887..265Y}) wherein the moving mesh is used as an approximation for the Lagrangian displacement field.

The moving-mesh construction is particularly well matched to the array-programming structure used above. As we will see, the deformation, metric construction, density deposition, elliptic solves, and particle update can all be written as local stencil or interpolation operations on regular arrays. This makes the method compatible with GPU execution and with reverse-mode automatic differentiation. The price is that the deformed mesh induces an anisotropic, spatially varying elliptic operator, so the FFT no longer diagonalises the Poisson equation. Non-uniform FFT methods can be used for irregular sampling problems \citep{lin2018python,2026arXiv260510678F}, but they do not remove the global spectral character of the solve and their effective bandwidth must still resolve the smallest local cell spacing.\footnote{More specifically, NUFFT does not remove the global nature of the spectral solve since the retained bandwidth must still resolve the smallest spatial scale anywhere in the domain. For a strongly adaptive moving mesh, this can make the number of modes comparable to a uniform grid at the finest local resolution, eroding the memory and scaling advantages of adaptivity. A geometric multigrid solve on the moving-mesh Laplacian instead acts on the adaptive degrees of freedom directly, recovering mesh-local scaling up to convergence factors.} We therefore use geometric multigrid as the natural elliptic solver for the moving geometry.

Below we describe the coordinate map, the induced metric, the curvilinear Poisson operator, and the coupled particle--mesh update. We then use the implementation to study cosmological structure formation and to demonstrate that the full moving-mesh forward model can be differentiated end to end. We provide a guide of symbols in this section in Table \ref{tab:symbols}.

\subsection{Coordinates, fields, and conventions}
\label{sec:coords}

Matter is represented by particles that carry computational (mesh)
coordinates $\xixi\in[0,n_g)^3$ together with a canonical momentum $\bm p$. The
computational domain is a periodic cube discretised on an $n_g^3$ grid. The grid
is deformed by a scalar deformation potential $\psi(\xixi)$, and physical
positions are recovered from computational ones through a potential--flow
(gradient) map,
\begin{equation}
  \xx(\xixi) \;=\; \xixi \;+\; \nabla_{\!\xixi}\,\psi(\xixi).
  \label{eq:map}
\end{equation}
The displacement of each cell from its undeformed location is therefore the
gradient $\nabla\psi$, and the entire deformed geometry is determined by the
single field $\psi$. We work throughout in cell units, taking the grid spacing to
be unity; we restore a physical spacing $\dd x$ as a matter of dimensional
bookkeeping and indicated where relevant. All fields are periodic, and the
discrete gradient is the second--order central difference
\begin{equation}
  (\nabla u)_\alpha \;=\; \frac{u_{+\bm e_\alpha}-u_{-\bm e_\alpha}}{2\,\dd x}.
\end{equation}
Because the deformation is a pure gradient, only the symmetric part of the
Jacobian is dynamical, and the geometry collapses onto the second derivatives of
$\psi$, as we now show.

\begin{table}
\centering
\begin{tabular}{ll}
\toprule
Symbol & Meaning \\
\midrule
$\psi$ & deformation potential (zero spatial mean) \\
$\dot\psi$ & deformation--rate potential, $\partial_\tau\psi$ \\
$\Amat$ & triad / Jacobian of the map, $\partial\xx/\partial\xixi$ \\
$\gmat=\Amat^\top\Amat$ & induced metric \\
$\sqrt{g}=\det\Amat$ & local volume element (physical/computational) \\
$\rho$ & matter density per unit physical volume \\
$\phi$ & gravitational potential \\
$\kappa$ & coupling controlling how strongly the mesh tracks mass \\
$\tau,\ a,\ E(a)$ & conformal time, scale factor, $H(a)/H_0$ \\
\bottomrule
\end{tabular}
\caption{Principal symbols used throughout the moving mesh solver.}\label{tab:symbols}
\end{table}

\subsection{Mesh distortion and the induced geometry}
\label{sec:geom}

Differentiating the map \eqref{eq:map} gives the Jacobian, which we call the
triad. Because the displacement is the gradient of a scalar, the triad is
symmetric and reduces to the identity plus the Hessian of the potential,
\begin{equation}
  \Amat(\xixi) \;=\; \mathbb{I} + \Hess\psi,
  \qquad
  \Amat_{\alpha\beta} \;=\; \delta_{\alpha\beta} + \partial_\alpha\partial_\beta\psi .
  \label{eq:triad}
\end{equation}
The second derivatives are evaluated with standard second--order central
stencils,
\begin{align}
  \partial_\alpha^2\psi
    &= \frac{\psi_{+\bm e_\alpha} - 2\psi + \psi_{-\bm e_\alpha}}{\dd x^2},\\
  \partial_\alpha\partial_\beta\psi
    &= \frac{\psi_{+\bm e_\alpha+\bm e_\beta} - \psi_{+\bm e_\alpha-\bm e_\beta}
           - \psi_{-\bm e_\alpha+\bm e_\beta} + \psi_{-\bm e_\alpha-\bm e_\beta}}{4\,\dd x^2}
      \quad(\alpha\neq\beta).
\end{align}
The metric of the deformed coordinates, its inverse, and the Jacobian determinant
follow algebraically,
\begin{equation}
  \gmat = \Amat^\top\Amat,\qquad
  \gmat^{-1} = \Amat^{-1}\Amat^{-\top},\qquad
  \sqrt{g} = \det\Amat .
  \label{eq:metric}
\end{equation}
The determinant $\sqrt{g}$ is the ratio of physical to computational cell volume
and is the central diagnostic of the method. Where the mesh contracts onto a
structure, $\sqrt{g}<1$, the local cell shrinks, and the effective resolution
rises in inverse proportion. Conversely $\sqrt{g}>1$ in evacuated regions, where
resolution is willingly sacrificed. A physically admissible mesh is one for which
the map \eqref{eq:map} remains invertible and unfolded. A positive volume element,
$\sqrt{g}=\det\Amat>0$, is necessary but not sufficient, since a positive
determinant also admits a pair of negative eigenvalues. The sharp condition is that
all eigenvalues of the symmetric triad remain positive, $\lambda_{\min}(\Amat)>0$,
which is precisely the quantity the compression limiter monitors. Maintaining this
condition under gravitational collapse is the central numerical challenge and the
reason for the limiters explored in  Appendix~\ref{sec:limiters}.

Particles are assigned to the grid by cloud--in--cell interpolation in
computational coordinates, producing a mesh mass field $m(\xixi)$. The
quantity required by the field equations is the density per unit physical
volume, obtained by dividing out the local volume element,
\begin{equation}
  \rho(\xixi) \;=\; \frac{m(\xixi)}{\sqrt{g}(\xixi)} .
  \label{eq:rho}
\end{equation}
This normalisation guarantees that the total mass is conserved on the deformed
mesh, in the sense that the volume--weighted sum reproduces the particle mass,
$\sum_{\xixi}\rho\,\sqrt{g}=M_{\rm tot}$. It is this $\sqrt{g}$ weighting that
makes the otherwise Eulerian deposition consistent with the moving geometry.

\subsection{The curvilinear multigrid Poisson solver}
\label{sec:mg}

The deformation introduces a spatially varying metric, so the elliptic problems
for gravity and for the mesh motion are not the flat Laplace equation but its
curvilinear (Laplace--Beltrami) generalisation,
\begin{equation}
  \mathcal{L}_\psi[u] \;\equiv\;
  \frac{1}{\sqrt{g}}\,\partial_\alpha\!\big(\sqrt{g}\,g^{\alpha\beta}\,\partial_\beta u\big)
  \;=\; f,
  \label{eq:LB}
\end{equation}
with periodic boundary conditions. For a fixed deformation field $\psi$, we denote
the corresponding discrete operator by $A_\psi$ defined such that,
\begin{equation}
  A_\psi u = f, \qquad r = f - A_\psi u .
  \label{eq:curv_linsys}
\end{equation}
The same operator, with the same geometry coefficients, serves both the
gravitational solve and the mesh--motion solve; only the source term differs.
Because $g^{\alpha\beta}$ varies from cell to cell, fast Fourier transforms no
longer diagonalise the operator and an iterative solver is required.

The multigrid cycle is the variable-coefficient analogue of the flat V-cycle in
\S\ref{sec:geommg}. After smoothing on the current level, the residual
$r^h=f^h-A_{\psi,h}u^h$ is restricted to the next coarser mesh, while the
deformation potential $\psi$ is restricted as well. The coarse operator is then
rebuilt from the coarse triad, metric, and volume element, yielding the error
equation
\begin{equation}
  A_{\psi,2h} e^{2h} = r^{2h} .
  \label{eq:curv_error}
\end{equation}
The prolongated correction updates the fine solution, $u^h\leftarrow u^h+P e^{2h}$,
and post-smoothing removes the high-frequency error introduced by interpolation.
This rediscretised hierarchy is slightly more expensive than the flat seven-point
case, but it preserves the conservative finite-volume form of the curvilinear
operator on each level and remains compatible with nearest-neighbor halo
communication. In the undeformed limit, $\psi=0$, $\sqrt{g}=1$, and
$g^{\alpha\beta}=\delta^{\alpha\beta}$, the hierarchy reduces exactly to the
standard periodic multigrid Poisson solver described above. We note that, for the variable-coefficient Laplace-Beltrami problem, we find the Jacobi smoothing described in \citet{2025arXiv251213403H} outperforms the Chebyshev smoothing used for the static mesh cases in Sec.~\ref{sec:geommg}.

\subsubsection{Conservative finite--volume discretisation}
The operator \eqref{eq:LB} is discretised in flux (divergence) form so as to
preserve its symmetry and its semidefiniteness. Defining the symmetric
coefficient field $\mathsf{B}^{\alpha\beta}=\sqrt{g}\,g^{\alpha\beta}$, the flux
$\bm F=\mathsf{B}\,\nabla u$ is evaluated on cell faces. The component of the flux
normal to a face uses the compact two--point difference across that face,
\begin{equation}
  (\partial_\alpha u)_{\alpha\text{--face}} = \frac{u_{+\bm e_\alpha}-u}{\dd x},
\end{equation}
whereas the tangential derivatives are formed by central differences at cell
centres and then averaged onto the face. The operator is the discrete divergence
of the face fluxes, divided by the local volume element,
\begin{equation}
  \mathcal{L}_\psi[u] \;=\; \frac{1}{\sqrt{g}\,\dd x}\,
  \Big[(F^x_{i+\frac12}-F^x_{i-\frac12}) + (\text{$y$ term}) + (\text{$z$ term})\Big].
\end{equation}
This construction is symmetric and negative--semidefinite, and in the undeformed
limit it reduces exactly to the standard seven--point Laplacian, whose Fourier
symbol is $\big(2\cos k_x+2\cos k_y+2\cos k_z-6\big)/\dd x^2$. In that flat limit
(used whenever the mesh is held fixed) the solve is performed directly and
exactly by FFT.

\begin{figure*}
    \centering
    \includegraphics[width=0.99\linewidth]{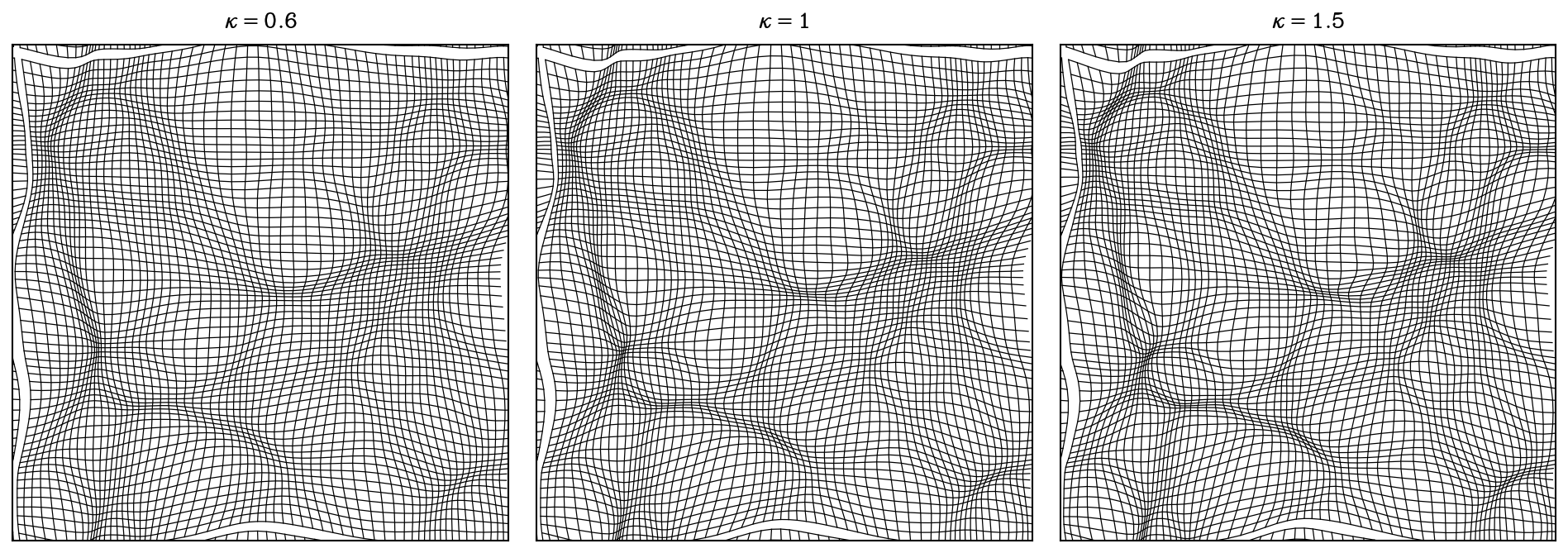}

    \caption{Effect of the mesh-response parameter $\kappa$ on the moving-mesh deformation at $z=0$. Each panel shows a two-dimensional slice through the deformed computational grid for a simulation run from the same initial condition but varying $\kappa$, with larger $\kappa$ driving stronger contraction toward filaments and halo-like peaks. Moderate values increase force resolution in overdense regions while maintaining a smooth, invertible map. Larger values of $\kappa$ require stronger limiter action to prevent folded or highly anisotropic cells.}
    \label{fig:kappa_demo}
\end{figure*}
\subsubsection{Solvability and gauge}

\begin{figure}
    \centering
    \includegraphics[width=0.99\linewidth]{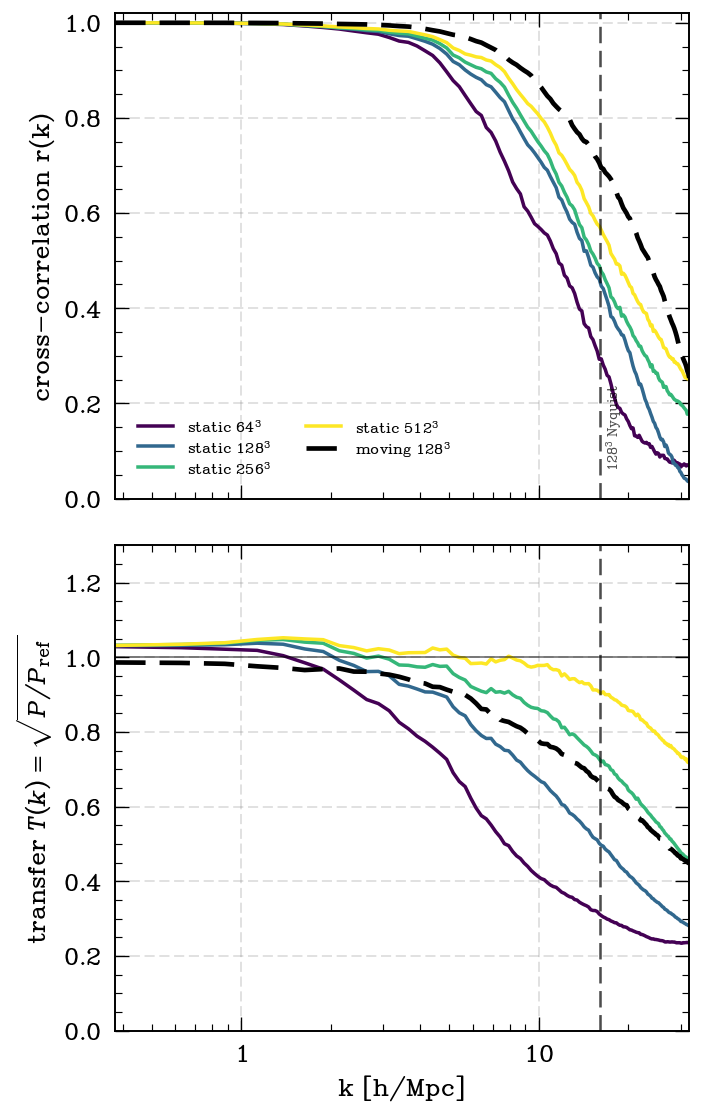}
    
    \caption{Clustering accuracy of static and moving-mesh PM calculations against the high-resolution reference at constant $256^3$ particle number. The upper panel shows the cross-correlation coefficient $r(k)$, and the lower panel shows the transfer function $T(k)=\sqrt{P/P_{\rm ref}}$. Increasing the static mesh resolution improves agreement, while the moving-mesh calculation recovers substantially more small-scale correlation than a static mesh with the same nominal number of cells.}
    \label{fig:cv0_static_ladder}
\end{figure}
\begin{figure}

 \centering
    \includegraphics[width=0.99\linewidth]{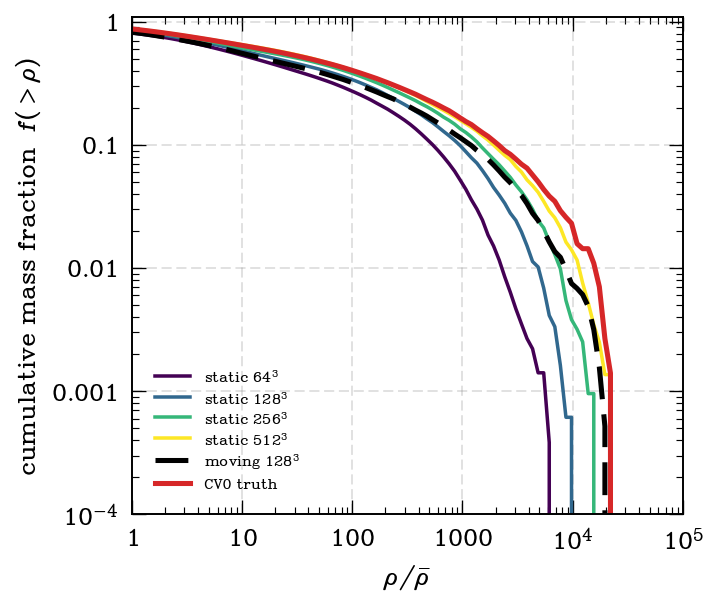}
    \caption{Cumulative mass fraction above density threshold for static PM meshes and the moving-mesh calculation. The moving mesh produces a higher-density tail than a static mesh of comparable nominal resolution, bringing the one-point distribution closer to the CV0 truth in collapsed regions. This is the real-space counterpart of the improved small-scale correlation shown in Fig.~\ref{fig:cv0_static_ladder}.}
    \label{fig:cv0_pdf}
\end{figure}
On a periodic domain the operator Eq.~\ref{eq:LB} is singular since constants lie in its kernel, and a solution exists only if the source is orthogonal to that kernel.
For the curved operator the correct compatibility condition is the
volume--weighted zero mean, which we enforce by projection,
\begin{equation}
  f \;\longleftarrow\; f - \frac{\sum_{\xixi} f\,\sqrt{g}}{\sum_{\xixi}\sqrt{g}},
  \qquad\Longrightarrow\qquad
  \sum_{\xixi} f\,\sqrt{g}=0 .
  \label{eq:solvability}
\end{equation}
The undetermined additive constant in the solution is fixed by removing its mean,
which is the natural gauge for a periodic potential.

\subsubsection{Cycle schedule and warm starting}
\label{sec:mgschedule}
In practice the hierarchy uses $L=\log_2 n_g-1$ levels with $2$
weighted--Jacobi sweeps for pre-- and post--smoothing, and we take a V--cycle
(one coarse--grid correction per level). For the range of mesh distortion
reached in these calculations a V--cycle is sufficient. A W--cycle (i.e. partially refining and then recoarsening)
performs substantially more coarse--grid work per iteration (of order $2^L$ rather
than $L$ smoother applications) for no measurable improvement in the recovered
field for the examples tried. We note this is different from the conclusion reached in \textsc{DiffHydro}, perhaps reflective of the different geometries and matter-distributions considered.

\begin{figure}
    \centering
    \includegraphics[width=0.99\linewidth]{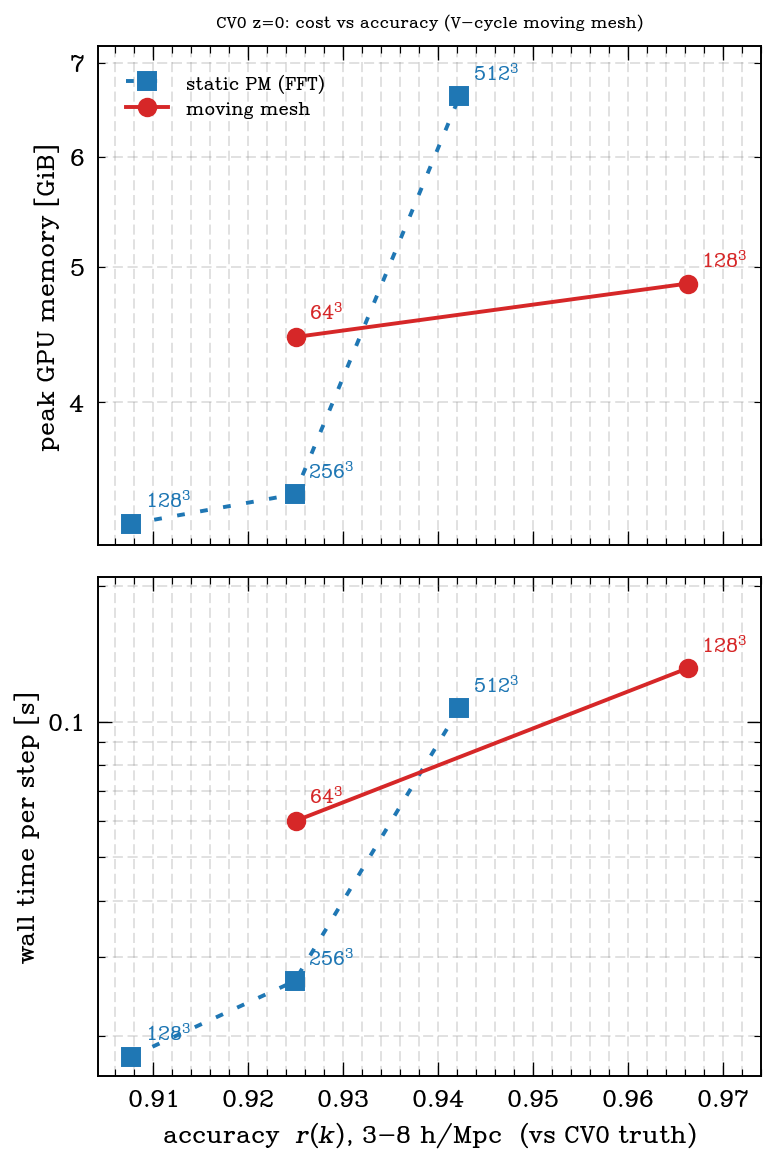}
    
    \caption{Cost--accuracy trade-off for static and moving-mesh PM calculations for fixed particle number $N=256^3$. Accuracy is summarized by the cross-correlation coefficient averaged over $k=3$--$8\,h\,\mathrm{Mpc}^{-1}$ relative to the CV0 truth, while the vertical axes show peak GPU memory and wall time per step. The moving mesh improves small-scale accuracy at fixed cell count by concentrating resolution into dense regions, but the curvilinear metric construction and multigrid solves add per-step overhead relative to a static FFT-based PM update. We note that due to dynamic allocation strategies of \textsc{JAX}, it is difficult to precisely benchmark memory usage.}
    \label{fig:cost_accuracy}
\end{figure}

We again use the warm-start procedure described in Sec.~\ref{sec:warmstart} to accelerate the convergence of our multigrid procedure. The gain is largest at late times, when the operator is most strongly deformed and a cold (zero) initial guess is worst. With warm starting, of order six V--cycles per solve converge the potential to the accuracy required for the force. Together with the combined--kick ordering above, this keeps the per--step cost of the moving mesh close to that of a static PM step at the resolution it replaces (Fig.~\ref{fig:cost_accuracy}).

\subsection{Evolution of the mesh: the deformation--rate potential}
\label{sec:defp}

The mesh is made to follow the matter by evolving the deformation potential along
a velocity that points down the density gradient. This velocity is itself the
gradient of a second scalar potential, the deformation--rate potential
$\dot\psi\equiv\partial_\tau\psi$, which is obtained from an elliptic solve whose
source is the local departure of the mass per computational cell
$m\equiv\sqrt{g}\,\rho$ from a uniform target value $m_{\rm init}$,
\begin{equation}
  \mathcal{L}_\psi[\dot\psi] \;=\; \kappa\,\big(m_{\rm init}-\sqrt{g}\,\rho\big),
  \label{eq:calcdefp}
\end{equation}
solved with the same curvilinear operator \eqref{eq:LB}. Driving the mesh toward
equal mass per cell, rather than equal density, is the natural quasi-Lagrangian
criterion \citep{1995ApJS..100..269P} and reduces to the density form only on a
uniform mesh ($\sqrt{g}=1$). The right--hand side is positive where the cell holds
more than the target mass, so the resulting flow $\nabla\dot\psi$ advects the mesh
inward, contracting cells onto mass concentrations and dilating them in voids. The
dimensionless gain $\kappa$
controls how vigorously the mesh responds to density contrast. Small $\kappa$
yields a nearly static grid, while larger $\kappa$ produces stronger adaptive
motion (see Fig.~\ref{fig:kappa_demo}). The source is smoothed
before the solve to suppress particle shot noise, which would otherwise drive spurious small--scale mesh motion and/or overly aggressive distortion.

The deformation is advanced by an explicit step in conformal time, after which
its mean is removed to preserve the gauge,
\begin{equation}
  \psi^{n+1} \;=\; \mathcal{Z}\!\big[\psi^{n} + \dd\tau\,\dot\psi\big],
  \qquad \mathcal{Z}[u]=u-\langle u\rangle .
  \label{eq:defadvance}
\end{equation}
The rate of change of the deformation defines a mesh velocity that the
particles must be referred to when they drift,
\begin{equation}
  \bm v_{\rm grid} \;=\; \nabla\!\left(\frac{\psi^{n+1}-\psi^{n}}{\dd\tau}\right)
  \;\approx\; \nabla\dot\psi .
  \label{eq:vgrid}
\end{equation}
Physically, $\bm v_{\rm grid}$ is the velocity of the computational cell centres
through physical space and the particle dynamics of Section~\ref{sec:force} are
written relative to this moving frame.

\subsection{Gravity and the particle force/drift update}
\label{sec:force}

\subsubsection{Gravitational potential and force}
The gravitational potential satisfies the curvilinear Poisson equation with the
projected density as its source,
\begin{equation}
  \mathcal{L}_\psi[\phi] \;=\; \mathcal{R}[\rho],
  \label{eq:gravity}
\end{equation}
where $\mathcal{R}$ denotes the solvability projection \eqref{eq:solvability}. The
gradient of $\phi$ is taken in computational coordinates and pulled back into
physical space through the inverse triad, yielding the physical acceleration
\begin{equation}
  \bm a \;=\; -\,\tfrac{3}{2}\,\Omega_m\;\Amat^{-1}\,\nabla\phi .
  \label{eq:accel}
\end{equation}
The prefactor $\tfrac32\Omega_m$ is the standard comoving Poisson normalisation
for cosmological structure growth, and the factor $\Amat^{-1}$ is what converts a
mesh--coordinate gradient into a physical force on the deformed grid. The mesh
velocity entering the drift is $\bm v_{\rm grid}$ from \eqref{eq:vgrid}.

\subsubsection{Particle update in the moving frame}
Particles carry the canonical momentum $\bm p = a\,\bm v$, where $\bm v$ is the
peculiar velocity. The grid fields $\bm a$, $\bm v_{\rm grid}$, and $\Amat^{-1}$
are interpolated to particle locations by cloud--in--cell, and the state is
advanced by a kick--drift--kick (leapfrog) sequence. Writing $\bm a(\xx)$ for the
acceleration \eqref{eq:accel} evaluated at the particle positions, an opening and a
closing half--kick bracket a drift in the moving frame,
\begin{align}
  \bm p^{\,n+1/2} &= \bm p^{\,n}
    + \tfrac12\,\bm a(\xx^{\,n})\,\dd\tau, \hspace{10pt}
  \bm v = \bm p^{\,n+1/2}/a_{\rm mid},
  \label{eq:kick}\\[2pt]
  \xixi^{\,n+1} &=
    \big(\xixi^{\,n}
    + \dd\tau\,\Amat^{-1}(\bm v - \bm v_{\rm grid})\big)\bmod n_g,
  \label{eq:drift}\\[2pt]
  \bm p^{\,n+1} &= \bm p^{\,n+1/2}
    + \tfrac12\,\bm a(\xx^{\,n+1})\,\dd\tau.
  \label{eq:kick2}
\end{align}
The opening force $\bm a(\xx^{\,n})$ is the closing force carried over from the
previous step (the combined kick), so within a step only the closing force
$\bm a(\xx^{\,n+1})$ (evaluated at the drifted positions on the advanced mesh
$\psi^{n+1}$) requires a fresh gravity solve.\footnote{The drift \eqref{eq:drift}
is written to leading order in $\dd\tau$. The implementation uses a second--order
expansion of the moving--frame displacement that additionally accounts for the
variation of the inverse triad over the step, $\dot{\bm b}=-\Amat^{-1}\dot\Amat\,
\Amat^{-1}$, i.e. the pull--back operator is $\Amat^{-1}\dd\tau+\tfrac12\dot{\bm b}\,
\dd\tau^2$.}
Equation \eqref{eq:drift} is the defining feature of the moving--mesh scheme. The
velocity of a particle in computational coordinates is its physical
peculiar velocity minus the velocity of the local cell, pulled back by the
inverse triad. When the mesh is frozen (i.e. $\dot\psi=0$,
$\bm v_{\rm grid}=0$, and $\Amat=\mathbb{I}$) the update reduces exactly to the
ordinary particle-mesh kick and drift. The moving-frame correction is therefore a strict generalization
that vanishes wherever the grid is undeformed.

\subsubsection{Cosmological time integration}
The integration is carried out in the scale factor with a uniform increment
$\dd a$, from which the conformal time--step follows as
$\dd\tau = \dd a/(a^2 E(a))$ with $E(a)=H(a)/H_0$. The sequence of scale factors
and corresponding steps is fixed in advance, and the per--step operations of the
following section are applied in a kick--drift--kick (leapfrog) ordering, which is
time--reversible in the gravity--only limit and second--order accurate in the
step.

\subsection{A complete time step}
\label{sec:step}
Collecting the pieces, one update of the coupled particle--mesh system proceeds as
follows:
\begin{enumerate}[label=\arabic*.,leftmargin=*,align=left]
  \item Deposit the particles and form the physical density $\rho=m/\sqrt{g}$ on the current mesh, Eq.~\eqref{eq:rho}.
  \item Solve for the mesh motion, $\mathcal{L}_\psi[\dot\psi]=\kappa(m_{\rm init}-\sqrt{g}\,\rho)$,
        applying the source--side and repulsive limiters,
        Eqs.~\eqref{eq:calcdefp},~\eqref{eq:rhslim},~\eqref{eq:repulsive}; advance the
        deformation $\psi^{n+1}=\mathcal{Z}[\psi+\dd\tau\,\dot\psi]$, Eq.~\eqref{eq:defadvance},
        and form the mesh velocity $\bm v_{\rm grid}$, Eq.~\eqref{eq:vgrid}.
  \item Apply the opening half--kick with the force carried from the previous step's
        closing solve (bootstrapped by one extra gravity solve on the first step only),
        Eq.~\eqref{eq:kick}.
  \item Drift the particles in the moving frame onto the advanced mesh $\psi^{n+1}$,
        Eq.~\eqref{eq:drift}.
  \item Re--deposit at the drifted positions and solve for gravity on the advanced mesh,
        $\mathcal{L}_\psi[\phi]=\mathcal{R}[\rho]$, Eq.~\eqref{eq:gravity} (curvilinear
        multigrid, or FFT in the frozen--mesh limit); apply the closing half--kick,
        Eqs.~\eqref{eq:accel},~\eqref{eq:kick2}, and retain this force to open the next step.
\end{enumerate}
The cost per step is dominated by the two elliptic solves,  one gravitational and
one for the mesh motion. Keeping the count at two relies on the combined--kick
ordering. A naive kick--drift--kick evaluates gravity twice per step (at the opening
and closing positions), but reusing the previous step's closing force as the current
opening force removes the start--of--step solve. Each multigrid solve scales linearly
with the number of cells, and the geometric overhead of forming the triad, metric,
and limiters is a small constant factor on top of a standard particle--mesh step.

\subsection{Moving-mesh accuracy tests}
\label{sec:mm_results}

We use the ``CV0" CAMELS dark matter \citep{2020arXiv201000619V} simulation evolved with the AREPO code \citep{2020AREPO} as a high-resolution reference to isolate the numerical effect of the moving mesh. The reference simulation was performed with $512^3$ particles in a $25$ $h^{-1}$Mpc side-length box at a fiducial cosmology of $\Omega_m=0.3$ and $\sigma_8 = 0.8$. The goal of this comparison is not to claim that a deformed mesh replaces a uniformly high-resolution calculation, but to quantify how much small-scale information can be recovered at fixed nominal cell count when the force resolution is redistributed toward collapsed regions. We therefore compare against a ladder of static PM calculations. We use the standard FFT implementation (rather than multi-grid) as we perform all calculations on one GPU where FFT is still optimal. This is a conservative baseline since the static solver has a simpler particle update, an exact FFT Poisson solve on the chosen mesh, and no metric-construction overhead.

As expected, increasing the resolution of the static PM mesh improves both quantities monotonically (Fig.~\ref{fig:cv0_static_ladder}). The moving-mesh calculation follows the same large-scale structure as the static run, but retains substantially more correlation on nonlinear scales at the same nominal number of cells. The potential-flow deformation moves cells into the filaments and halo-like peaks where force errors dominate the final density field.

The same behaviour is visible in the one-point statistics. Figure~\ref{fig:cv0_pdf} shows the cumulative mass fraction above a density threshold. Coarse static PM runs broaden collapsed regions and therefore suppress the high-density tail. The moving mesh restores part of this tail by allocating smaller cells to the overdense regions, bringing the density distribution closer to the CV0 reference in the regime most affected by force softening. This agreement is not perfect, and it is not expected to be: the moving mesh remains a low-dimensional, potential-flow deformation of a Cartesian grid, not a fully adaptive refinement hierarchy.

The cost of the method is shown in Fig.~\ref{fig:cost_accuracy}. Relative to static PM at the same cell count, the moving mesh adds metric construction, source limiting, a mesh-motion solve, and a variable-coefficient gravitational solve. This method is therefore most useful in the regime where small-scale accuracy is most important. In that regime, using the available cells more effectively can recover part of the high-$k$ phase information that would otherwise require a much larger uniform mesh. We view the moving mesh as occupying an intermediate position between cheap fixed-grid PM and substantially more expensive TreePM \citep{2020AREPO}, adaptive mesh refinement (i.e. \citet{2002A&A...385..337T}), or very high-resolution static calculations.

For the purposes of this work, the main result is that the curvilinear multigrid machinery produces the expected numerical behaviour. It improves small-scale clustering relative to a static mesh of comparable size, preserves the large-scale phases, and does so using only operations that remain compatible with the JAX differentiation and compilation model. This sets up the reconstruction experiment below, where the same adaptive force model is placed inside a gradient-based optimization loop.
\subsection{Differentiability of the forward map}
\label{app:ad}

A primary motivation for implementing the moving-mesh algorithm in JAX is that the
adaptive force calculation can be embedded, unchanged, inside the same differentiable
workflows as the static particle--mesh (PM) solver (i.e. standard \textsc{JaxPM} or \textsc{PMWD}). The entire evolution (deposition,
mesh-geometry construction, the curvilinear elliptic solves, the mesh-regularity
limiters, the particle kick and drift, and the final density readout) is expressed as a
composition of JAX primitives, so a single reverse-mode pass returns the gradient of any
scalar functional of the final state with respect to the initial conditions or the model
parameters. In this section we demonstrate that capability with a moving-mesh reconstruction
experiment, in the same spirit as the reconstruction examples in \textsc{diffhydro}. The
experiment is designed as a numerical stress test of the forward model and its adjoint
under strong mesh deformation, rather than as a complete cosmological inference analysis.
The target, loss, and optimization schedule are chosen to expose the numerical behaviour
of the method, and a production field-level inference pipeline would additionally require
a noise model, cosmological priors, nuisance parameters, and posterior validation, all of
which are beyond the scope of this demonstration \citep{2023arXiv230709504B,2025JCAP...12..039S}.

\subsubsection{Limiters in the reverse pass.}
Most operations in the forward map are smooth array primitives and can be differentiated
directly. The main exception is the family of mesh-regularity limiters described in
Appendix~\ref{sec:limiters}, which guard against mesh folding through clips and threshold
switches. These limiters are numerical safeguards, not physical interactions, and their
non-smooth branches are not useful search directions for gradient-based inference. We
therefore evaluate them in the forward pass but hold the limiter modification fixed in the
reverse pass using a straight-through construction. If $f_{\rm phys}=\kappa(\rho_{\rm
init}-\rho)$ is the smooth mesh-motion source and $f_{\rm lim}$ is the source after
limiting, we write
\begin{equation}
  f = f_{\rm phys} + \texttt{stop\_gradient}(f_{\rm lim}-f_{\rm phys}) ,
  \label{eq:straight_through_limiter}
\end{equation}
so that the forward value is the fully limited source while the gradient follows the
smooth physical source and the elliptic solves. The same treatment is applied to the
repulsive correction of Eq.~\eqref{eq:repulsive}. This choice keeps the adjoint well
behaved while preserving exactly the forward dynamics that maintain a valid, non-folded
mesh.

\begin{figure*}
    \centering
    \includegraphics[width=0.90\linewidth]{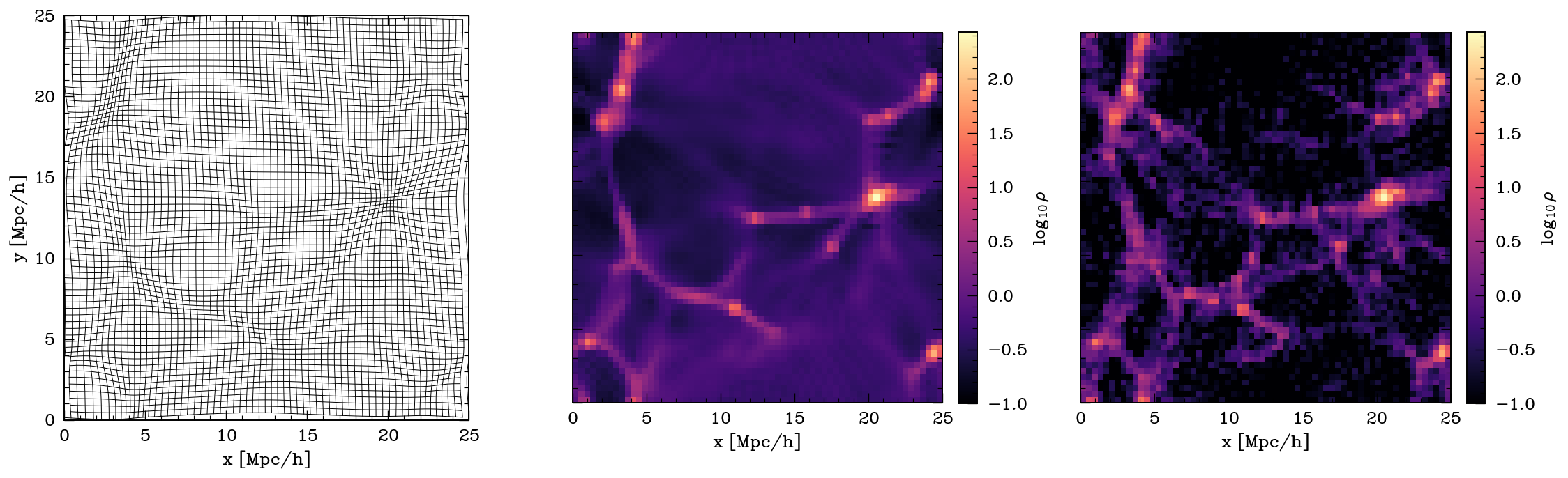}
    \caption{Differentiable moving-mesh reconstruction in a $25\,h^{-1}{\rm Mpc}$ slice,
    shown as a projected five-cell slab. \textit{Left:} the late time mesh for the optimized field,
    which contracts around the filaments and halo-like peaks that dominate the nonlinear
    density field; in the densest regions the cells are compressed to roughly a sixth of
    their reference volume ($\sqrt{g}_{\min}\!\approx\!0.15$). \textit{Middle:} the
    reconstructed logarithmic density obtained by differentiating through the moving-mesh
    particle--mesh evolution. \textit{Right:} the target $N$-body density on the same
    slice. The reconstruction reproduces the cosmic-web morphology and the positions and
    amplitudes of the dominant peaks.} 
    \label{fig:recon_density}
\end{figure*}
\begin{figure}
    \centering
    \includegraphics[width=0.99\linewidth]{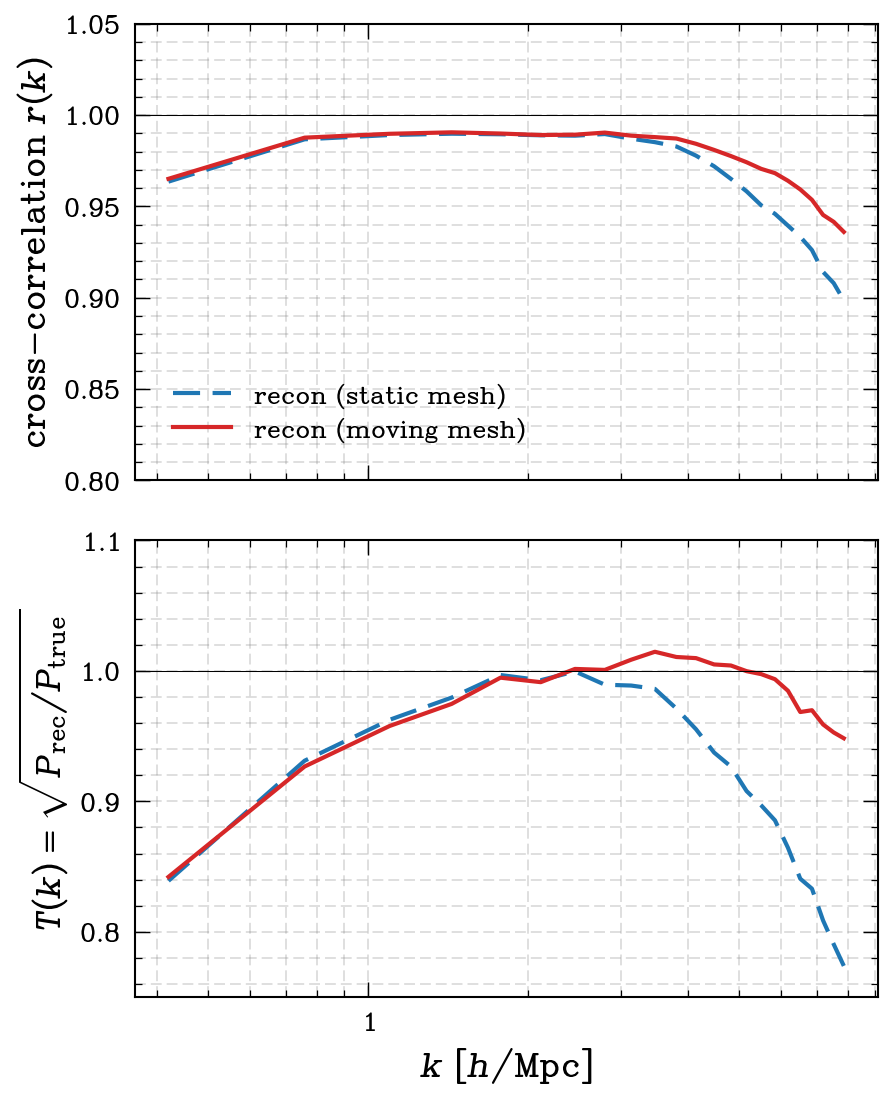}
    \caption{Fourier-space diagnostics for static-mesh (i.e. \textsc{JaxPM}) and moving-mesh differentiable
    reconstructions of the same target field. \textit{Upper:} the cross-correlation
    coefficient $r(k)$ between the reconstructed and target $z=0$ density fields.
    \textit{Lower:} the transfer function $T(k)=\sqrt{P_{\rm rec}/P_{\rm true}}$. The two
    reconstructions agree on large and intermediate scales and separate toward the grid
    Nyquist frequency, where the moving mesh retains $T\!\approx\!0.95$ and a higher
    correlation while the static mesh is suppressed to $T\!\approx\!0.78$.} 
    \label{fig:recon_stats}
\end{figure}

\subsubsection{Adjoint of the curvilinear elliptic solves.}
The gravitational potential and the mesh-motion (deformation) potential are each obtained
from a curvilinear Poisson problem, solved with a fixed number of geometric multigrid
cycles. A curvilinear solve can be differentiated by unrolling the cycle sequence through
the reverse pass, but this stores the entire solver trajectory and, at a finite cycle
count, back-propagates through an only partially converged iteration. We instead attach an
implicit (adjoint-state) vector--Jacobian product to the linear solve. Writing the
discrete curvilinear operator as
\begin{equation}
  A(\mathbf{s}) = W(\mathbf{s})^{-1}\,M(\mathbf{s}), \qquad W=\mathrm{diag}(\sqrt{g}),
  \label{eq:curv_operator}
\end{equation}
where $\mathbf{s}$ denotes the mesh deformation, $\sqrt{g}$ the local cell volume (the
determinant of the metric), and $M=M^{\!\top}$ the symmetric flux-divergence operator, the
solve returns $\phi = A^{-1}F$ for a source $F$. For a cotangent $\bar{\phi}$ carried on
the solution, reverse-mode differentiation of the implicit relation
$A(\mathbf{s})\,\phi=F$ gives
\begin{align}
  \bar{F} &= A^{-\top}\bar{\phi} \;=\; \sqrt{g}\, A^{-1}\!\big(\bar{\phi}/\sqrt{g}\big),
  \label{eq:adj_rhs}\\[2pt]
  \bar{\mathbf{s}} &= -\big[\,\partial_{\mathbf{s}}\big(A(\mathbf{s})\,\phi\big)\big]^{\!\top}\,\bar{F},
  \label{eq:adj_op}
\end{align}
with all multiplication/division by $\sqrt{g}$ understood to be pointwise. Analogous to the static mesh case discussed in \citet{2025arXiv251213403H}, Equation~\eqref{eq:adj_rhs} is the \emph{same} multigrid solver applied to the
$\sqrt{g}$-conjugated cotangent\footnote{This is a consequence of $M$ being self-adjoint in the
$\sqrt{g}$-weighted inner product rather than the Euclidean one.} and
Eq.~\eqref{eq:adj_op} is the sensitivity of the operator to the mesh geometry, evaluated
at the converged solution $\phi$ and contracted with the adjoint field. The latter reduces
to a single vector--Jacobian product of the (inexpensive) operator application and adds
negligible cost relative to the forward solve. 

Discarding both the $\sqrt{g}$ conjugation
in Eq.~\eqref{eq:adj_rhs} and the operator term in Eq.~\eqref{eq:adj_op} (i.e. simply
re-running the forward solver on $\bar{\phi}$) yields an adjoint that is exact only for an
undeformed mesh ($\sqrt{g}\equiv1$). The error grows with the mesh compression and, left
uncorrected, systematically biases the reconstruction gradient and caps the deformation
amplitude that gradient descent can reach. With both terms retained, a directional
finite-difference test of the complete rollout agrees with the automatic-differentiation
gradient to better than one percent even when cells in the densest regions are compressed
to $\sim\!15\%$ of their reference volume. Because both elliptic solves share the same
curvilinear operator, a single adjoint implementation serves the gravitational and the
mesh-motion potentials. 


\subsubsection{Reconstruction experiment.}
The optimization variable is the initial linear density field that seeds the particle
distribution through first-order Lagrangian perturbation theory. The forward model
advances this field with the moving-mesh PM integrator to $z=0$ and deposits the particles
onto a fixed Eulerian grid, and the loss is a mean-square difference between the resulting
density and a target density field. We take as the target the $z=0$ density of a full
$N$-body simulation from the CAMELS suite, deposited on the same grid, so that the
recoverable accuracy is bounded both by the fidelity of the moving-mesh PM relative to the
reference $N$-body dynamics and by the invertibility of the forward map. We minimize the
residual plus a Gaussian phase prior with a guarded Adam optimizer, warm-started from a static-mesh ($\kappa=0$)
reconstruction and annealed along two variables: the mesh-adaptivity parameter $\kappa$
is ramped from zero to its target value so that the geometry deforms gradually, and the
loss is filtered with a Fourier cutoff swept from large scales to the grid Nyquist
frequency. This coarse-to-fine schedule keeps the optimization on the smooth part of the
landscape while the mesh (and with it the elliptic adjoint) is driven into the strongly deformed regime. 

Figure~\ref{fig:recon_density} shows an optimized moving-mesh reconstruction in a $25\,h^{-1}{\rm Mpc}$ slice. The mesh contracts onto the same filaments and halo-like peaks that dominate the target field, allocating force resolution to the regions that dominate the loss while preserving the structured array layout required for JAX compilation and reverse-mode differentiation. The reconstructed density reproduces the large-scale web and recovers the highest density peak to within a few percent of the true value, where the corresponding static-mesh reconstruction undershoots the same peak by of order ten percent. The best global cross-correlation, $r\!\approx\!0.97$, is obtained at moderate deformation. The mesh can be driven substantially further (to
$\sqrt{g}_{\min}\!\approx\!0.15$ in the densest cells, comparable to the compression of the true nonlinear field) while still maintaining $r>0.95$, which is the configuration shown
in Fig.~\ref{fig:recon_density}. 

Figure~\ref{fig:recon_stats} provides the complementary Fourier-space diagnostics. The
moving- and static-mesh reconstructions are nearly indistinguishable on large and
intermediate scales, where both recover the same coherent displacement field, and they
diverge toward the grid Nyquist frequency. There the moving mesh holds the transfer
function $T(k)=\sqrt{P_{\rm rec}/P_{\rm true}}$ close to unity ($T\!\approx\!0.95$) and
retains a higher cross-correlation, whereas the static mesh is suppressed to
$T\!\approx\!0.78$ as it fails to represent the collapsed structure at fixed resolution. In
this sense the reconstruction experiment is an end-to-end test of the central claim of
this section: the adaptive force solver is not merely differentiable in a formal sense, but
supplies gradients that recover additional nonlinear, small-scale amplitude at fixed
nominal resolution. The transfer deficit in the lowest-$k$ bins ($T\!\approx\!0.85$) is
common to both meshes. It  is not removed by the adaptive geometry and reflects a genuine
information limit of reconstructing a nonlinearly evolved field in a small box, where the
few available large-scale modes are weakly constrained because gravitational mode-coupling
redistributes their initial power across a range of scales.

\section{Discussion and Conclusions}
\label{sec:disc}
In this work we have revisited geometric multigrid Poisson solvers for particle--mesh cosmology in the regime relevant for modern autodiff and GPU-native simulation workflows. FFT solvers provide an exact spectral inversion on the chosen grid, but they require global transposes and have a fixed communication pattern. Multigrid solvers use local stencil operations and halo exchange, and their accuracy can be tuned to the needs of the calculation. These properties become particularly valuable for distributed GPU runs, repeated time-stepped solves, and forward models that must be differentiated.

For the static FastPM application, we find that the relevant operating point is not a cold multigrid solve but a warm-started Chebyshev-smoothed V-cycle. In the large-mesh tests considered in this work, one or two warm cycles reduce both wall-clock cost and memory pressure relative to the FFT path while maintaining sufficient accuracy. The measured speedups are necessarily machine dependent, because distributed FFT performance depends strongly on the interconnect and library implementation, but the memory and communication advantages are more structural. The FFT-free evolution avoids the complex transposed scratch arrays required by the spectral force calculation.

The moving-mesh calculation addresses a different limitation of fixed-grid PM. In cosmological simulations, uniform meshes spend most of their cells in voids, while the largest force errors for nonlinear clustering occur in collapsed regions. The potential-flow deformation used here concentrates cells around filaments and halo-like peaks while preserving a regular array topology. This makes the method closer to an ordinary PM code than to a tree or fully unstructured adaptive method with deposition, geometry construction, elliptic solves, and particle updates remain vectorized array operations. The resulting Poisson equation is a variable-coefficient curvilinear problem, so the FFT is no longer the appropriate solver; geometric multigrid is instead the numerical component that makes the method practical.

The numerical tests show the expected trade-off between small scale accuracy and wall-clock time. At fixed nominal resolution, the moving mesh improves the small-scale cross-correlation, the transfer function, and the high-density tail relative to a static PM calculation. At the same time, the method adds metric construction, mesh-motion solves, limiter operations, and variable-coefficient multigrid cycles. It is therefore not necessarily a universal replacement for increasing the static mesh resolution. Rather, it occupies an intermediate regime where memory limits prevent a much finer uniform grid, the target observables are dominated by overdense regions, or one needs an adaptive force model that remains differentiable and compatible with accelerator compilation.

The current moving-mesh implementation uses limiter parameters that are numerical rather than physical, including the maximum compression, maximum skew, source smoothing, and mesh-response strength. These choices should be calibrated more systematically across redshift, cosmology, halo mass, and particle load. Finally, the reconstruction example is intentionally minimal. A realistic field-level inference application would need a proper likelihood, priors, nuisance parameters, longer-run adjoints, and validation that the approximate adaptive dynamics do not bias posterior constraints. As these methods can push to smaller scales for a given grid size, it becomes important to consider more nuanced forward models than simple bias implementations, such as halo based methods (e.g. \citet{2024MNRAS.529.2473H,2026MNRAS.tmp.1081H}) or environmental bias (e.g. \citet{2026A&A...709A.110R}).

Despite these caveats, the results demonstrate that GPU-native geometric multigrid provides a useful complement to FFT-based Poisson solving in differentiable cosmological simulations. On fixed meshes, warm-started Chebyshev multigrid reduces global communication and memory requirements while retaining a controllable level of field-level accuracy. On moving meshes, the same framework enables spatially varying force resolution without abandoning the regular array operations and accelerator-friendly programming model of particle--mesh methods. This combination may be particularly valuable for increasingly large simulations, for which the communication and memory costs of distributed FFTs become important, as well as for applications in which the scientifically relevant structures occupy only a fraction of the simulated volume.

A natural next step is to couple the moving-mesh gravity solver to hydrodynamics, as in \citet{1998ApJS..115...19P}. The coordinate transformation can be applied consistently to the conservation laws for mass, momentum, and energy, allowing both gravitational and hydrodynamic resolution to be concentrated in dynamically important regions. The original \citet{1998ApJS..115...19P} approach was subsequently used in cosmological cluster \citep{1999ApJ...525..554F} and Sunyaev--Zel'dovich simulations \citep{2002MNRAS.336.1256K}, where it provided competitive shock resolution and substantial dynamic range without requiring uniformly high spatial resolution. Its broader adoption was nevertheless limited by the fixed logical connectivity of the globally deformed mesh. Strongly nonlinear, shearing, and multiply collapsing flows could produce distorted cells, requiring smoothing and compression limits that reduced the effective adaptivity. Since those earlier works, adaptive mesh refinement, unstructured mesh (i.e. Voronoi cells), and mesh-less approaches have became dominant in the astrophysics literature \citep{2025arXiv250206954V}. 

Despite these limitations, the moving mesh approach is worth revisiting in a modern differentiable setting. As shown in Sec.~\ref{sec:mm_results}, fixed topology grid maps naturally onto accelerator-oriented array programming, its gravity and deformation equations can share a common multigrid infrastructure, and differentiability allows the deformation itself to be optimized for a specified downstream observable rather than prescribed solely through a density-based heuristic (similar to the flux-reconstruction solver-in-the-loop methods shown in \citet{2025arXiv251213403H}). This creates the possibility of allocating resolution according to the needs of a particular scientific application, while explicit regularization controls mesh distortion and numerical stability.


Together, these results suggest a path toward cosmological solvers that preserve the simplicity, composability, and accelerator efficiency of array-based particle--mesh methods while gaining some of the spatial flexibility of adaptive and tree-based approaches. Extending this framework to self-consistent hydrodynamics, jointly optimizing the coordinate deformation and numerical solver, and testing the resulting models against observable-level accuracy requirements will be important directions for future work.

\section*{Acknowledgements}

I would like to thank Wassim Kabalan, Ue-Li Pen, Yu Yu, Peter Behroozi, Kentaro Nagamine, Zarija Lukic, and Adrian Bayer for useful discussions and comments on earlier drafts. I would particularly like to thank Wassim Kabalan for help in implementing \textsc{jaxDecomp} and Ue-Li Pen for the Fortran moving mesh implementation. This research used resources of the National Energy Research Scientific Computing Center (NERSC), a Department of Energy User Facility. This work made use of AI tools, including ChatGPT and Claude-AI. Kavli IPMU was established by World Premier International Research Center Initiatives (WPI), MEXT, Japan. This work was performed in part at the Center for Data-Driven Discovery, Kavli IPMU (WPI). This project use AI tools (Claude Opus) to assist in analysis.

\section*{Data Availability Statement}

No new observational data were collected for this work. Development versions of the multigrid and moving-mesh implementations are available at \url{https://github.com/bhorowitz/MMMM} and \url{https://github.com/bhorowitz/JaxPM}; the final versions used in this study will be released upon publication.

\bibliographystyle{mnras}
\bibliography{example} 

@ARTICLE{2019ApJ...887..265Y,
       author = {{Yu}, Yu and {Zhu}, Hong-Ming},
        title = "{Nonlinear Reconstruction of the Velocity Field}",
      journal = {\apj},
         year = 2019,
        month = dec,
       volume = {887},
       number = {2},
          eid = {265},
        pages = {265},
          doi = {10.3847/1538-4357/ab5580},
archivePrefix = {arXiv},
       eprint = {1908.08217},
 primaryClass = {astro-ph.CO},
       adsurl = {https://ui.adsabs.harvard.edu/abs/2019ApJ...887..265Y}
}

@ARTICLE{2011SJSC...33.2864B,
       author = {{Baker}, Allison H. and {Falgout}, Robert D. and {Kolev}, Tzanio V. and {Yang}, Ulrike Meier},
        title = "{Multigrid Smoothers for Ultraparallel Computing}",
      journal = {SIAM Journal on Scientific Computing},
         year = 2011,
        month = jan,
       volume = {33},
       number = {5},
        pages = {2864-2887},
          doi = {10.1137/100798806},
       adsurl = {https://ui.adsabs.harvard.edu/abs/2011SJSC...33.2864B}
}

@ARTICLE{2003JCoPh.188..593A,
       author = {{Adams}, Mark and {Brezina}, Marian and {Hu}, Jonathan and {Tuminaro}, Ray},
        title = "{Parallel multigrid smoothing: polynomial versus Gauss-Seidel}",
      journal = {Journal of Computational Physics},
         year = 2003,
        month = jul,
       volume = {188},
       number = {2},
        pages = {593-610},
          doi = {10.1016/S0021-9991(03)00194-3},
       adsurl = {https://ui.adsabs.harvard.edu/abs/2003JCoPh.188..593A}
}

@ARTICLE{2025arXiv250206954V,
       author = {{Valentini}, Milena and {Dolag}, Klaus},
        title = "{Hydrodynamic methods and sub-resolution models for cosmological simulations}",
      journal = {arXiv e-prints},
         year = 2025,
        month = feb,
          eid = {arXiv:2502.06954},
        pages = {arXiv:2502.06954},
          doi = {10.48550/arXiv.2502.06954},
archivePrefix = {arXiv},
       eprint = {2502.06954},
 primaryClass = {astro-ph.CO},
       adsurl = {https://ui.adsabs.harvard.edu/abs/2025arXiv250206954V}
}

@ARTICLE{1999ApJ...525..554F,
       author = {{Frenk}, C.~S. and {White}, S.~D.~M. and {Bode}, P. and {Bond}, J.~R. and {Bryan}, G.~L. and {Cen}, R. and {Couchman}, H.~M.~P. and {Evrard}, A.~E. and {Gnedin}, N. and {Jenkins}, A. and {Khokhlov}, A.~M. and {Klypin}, A. and {Navarro}, J.~F. and {Norman}, M.~L. and {Ostriker}, J.~P. and {Owen}, J.~M. and {Pearce}, F.~R. and {Pen}, U.-L. and {Steinmetz}, M. and {Thomas}, P.~A. and {Villumsen}, J.~V. and {Wadsley}, J.~W. and {Warren}, M.~S. and {Xu}, G. and {Yepes}, G.},
        title = "{The Santa Barbara Cluster Comparison Project: A Comparison of Cosmological Hydrodynamics Solutions}",
      journal = {\apj},
         year = 1999,
        month = nov,
       volume = {525},
       number = {2},
        pages = {554-582},
          doi = {10.1086/307908},
archivePrefix = {arXiv},
       eprint = {astro-ph/9906160},
 primaryClass = {astro-ph},
       adsurl = {https://ui.adsabs.harvard.edu/abs/1999ApJ...525..554F}
}

@ARTICLE{2002MNRAS.336.1256K,
       author = {{Komatsu}, E. and {Seljak}, U.},
        title = "{The Sunyaev-Zel'dovich angular power spectrum as a probe of cosmological parameters}",
      journal = {\mnras},
         year = 2002,
        month = nov,
       volume = {336},
       number = {4},
        pages = {1256-1270},
          doi = {10.1046/j.1365-8711.2002.05889.x},
archivePrefix = {arXiv},
       eprint = {astro-ph/0205468},
 primaryClass = {astro-ph},
       adsurl = {https://ui.adsabs.harvard.edu/abs/2002MNRAS.336.1256K}
}

@ARTICLE{2024MNRAS.529.2473H,
       author = {{Horowitz}, Benjamin and {Hahn}, ChangHoon and {Lanusse}, Francois and {Modi}, Chirag and {Ferraro}, Simone},
        title = "{Differentiable stochastic halo occupation distribution}",
      journal = {\mnras},
         year = 2024,
        month = apr,
       volume = {529},
       number = {3},
        pages = {2473-2482},
          doi = {10.1093/mnras/stae350},
archivePrefix = {arXiv},
       eprint = {2211.03852},
 primaryClass = {astro-ph.CO},
       adsurl = {https://ui.adsabs.harvard.edu/abs/2024MNRAS.529.2473H}
}

@ARTICLE{2017ApJ...847..110Y,
       author = {{Yu}, Yu and {Zhu}, Hong-Ming and {Pen}, Ue-Li},
        title = "{Halo Nonlinear Reconstruction}",
      journal = {\apj},
         year = 2017,
        month = oct,
       volume = {847},
       number = {2},
          eid = {110},
        pages = {110},
          doi = {10.3847/1538-4357/aa89e7},
archivePrefix = {arXiv},
       eprint = {1703.08301},
 primaryClass = {astro-ph.CO},
       adsurl = {https://ui.adsabs.harvard.edu/abs/2017ApJ...847..110Y}
}

@ARTICLE{2018PhRvD..97d3502Z,
       author = {{Zhu}, Hong-Ming and {Yu}, Yu and {Pen}, Ue-Li},
        title = "{Nonlinear reconstruction of redshift space distortions}",
      journal = {\prd},
         year = 2018,
        month = feb,
       volume = {97},
       number = {4},
          eid = {043502},
        pages = {043502},
          doi = {10.1103/PhysRevD.97.043502},
archivePrefix = {arXiv},
       eprint = {1711.03218},
 primaryClass = {astro-ph.CO},
       adsurl = {https://ui.adsabs.harvard.edu/abs/2018PhRvD..97d3502Z}
}

@article{eastwood1974shaping,
  title={Shaping the force law in two-dimensional particle-mesh models},
  author={Eastwood, James W and Hockney, Roger Williams},
  journal={Journal of Computational Physics},
  volume={16},
  number={4},
  pages={342--359},
  year={1974},
  publisher={Elsevier}
}

@BOOK{1981csup.book.....H,
       author = {{Hockney}, R.~W. and {Eastwood}, J.~W.},
        title = "{Computer Simulation Using Particles}",
         year = 1981,
       adsurl = {https://ui.adsabs.harvard.edu/abs/1981csup.book.....H}
}

@ARTICLE{1985ApJS...57..241E,
       author = {{Efstathiou}, G. and {Davis}, M. and {White}, S.~D.~M. and {Frenk}, C.~S.},
        title = "{Numerical techniques for large cosmological N-body simulations}",
      journal = {\apjs},
         year = 1985,
        month = feb,
       volume = {57},
        pages = {241-260},
          doi = {10.1086/191003},
       adsurl = {https://ui.adsabs.harvard.edu/abs/1985ApJS...57..241E}
}

@ARTICLE{1997astro.ph.12217K,
       author = {{Klypin}, Anatoly and {Holtzman}, Jon},
        title = "{Particle-Mesh code for cosmological simulations}",
      journal = {arXiv e-prints},
         year = 1997,
        month = dec,
          eid = {astro-ph/9712217},
        pages = {astro-ph/9712217},
          doi = {10.48550/arXiv.astro-ph/9712217},
archivePrefix = {arXiv},
       eprint = {astro-ph/9712217},
 primaryClass = {astro-ph},
       adsurl = {https://ui.adsabs.harvard.edu/abs/1997astro.ph.12217K}
}

@ARTICLE{2020ApJS..249....4S,
       author = {{Stone}, James M. and {Tomida}, Kengo and {White}, Christopher J. and {Felker}, Kyle G.},
        title = "{The Athena++ Adaptive Mesh Refinement Framework: Design and Magnetohydrodynamic Solvers}",
      journal = {\apjs},
         year = 2020,
        month = jul,
       volume = {249},
       number = {1},
          eid = {4},
        pages = {4},
          doi = {10.3847/1538-4365/ab929b},
archivePrefix = {arXiv},
       eprint = {2005.06651},
 primaryClass = {astro-ph.IM},
       adsurl = {https://ui.adsabs.harvard.edu/abs/2020ApJS..249....4S}
}

@ARTICLE{2005NewA...10..393M,
       author = {{Merz}, Hugh and {Pen}, Ue-Li and {Trac}, Hy},
        title = "{Towards optimal parallel PM N-body codes: PMFAST}",
      journal = {\na},
         year = 2005,
        month = apr,
       volume = {10},
       number = {5},
        pages = {393-407},
          doi = {10.1016/j.newast.2005.02.001},
archivePrefix = {arXiv},
       eprint = {astro-ph/0402443},
 primaryClass = {astro-ph},
       adsurl = {https://ui.adsabs.harvard.edu/abs/2005NewA...10..393M}
}

@article{brandt1977multi,
  title={Multi-level adaptive solutions to boundary-value problems},
  author={Brandt, Achi},
  journal={Mathematics of computation},
  volume={31},
  number={138},
  pages={333--390},
  year={1977}
}

@article{fedorenko1962relaxation,
  title={A relaxation method for solving elliptic difference equations},
  author={Fedorenko, Radii Petrovich},
  journal={USSR Computational Mathematics and Mathematical Physics},
  volume={1},
  number={4},
  pages={1092--1096},
  year={1962},
  publisher={Elsevier}
}

@ARTICLE{1997ApJS..111...73K,
       author = {{Kravtsov}, Andrey V. and {Klypin}, Anatoly A. and {Khokhlov}, Alexei M.},
        title = "{Adaptive Refinement Tree: A New High-Resolution N-Body Code for Cosmological Simulations}",
      journal = {\apjs},
         year = 1997,
        month = jul,
       volume = {111},
       number = {1},
        pages = {73-94},
          doi = {10.1086/313015},
archivePrefix = {arXiv},
       eprint = {astro-ph/9701195},
 primaryClass = {astro-ph},
       adsurl = {https://ui.adsabs.harvard.edu/abs/1997ApJS..111...73K}
}

@ARTICLE{1995ApJS..100..269P,
       author = {{Pen}, Ue-Li},
        title = "{A Linear Moving Adaptive Particle-Mesh N-Body Algorithm}",
      journal = {\apjs},
         year = 1995,
        month = sep,
       volume = {100},
        pages = {269},
          doi = {10.1086/192219},
       adsurl = {https://ui.adsabs.harvard.edu/abs/1995ApJS..100..269P}
}

@INPROCEEDINGS{1998assp....3...34F,
       author = {{Frigo}, M. and {Johnson}, S.~G.},
        title = "{FFTW: an adaptive software architecture for the FFT}",
    booktitle = {Proceedings of the 1998 IEEE International Conference on Acoustics},
         year = 1998,
       volume = {3},
        month = jan,
          eid = {34},
        pages = {34},
          doi = {10.1109/ICASSP.1998.681704},
       adsurl = {https://ui.adsabs.harvard.edu/abs/1998assp....3...34F}
}

@ARTICLE{2005IEEEP..93..216F,
       author = {{Frigo}, Matteo and {Johnson}, Steven G.},
        title = "{The Design and Implementation of FFTW3}",
      journal = {IEEE Proceedings},
         year = 2005,
        month = jan,
       volume = {93},
       number = {2},
        pages = {216-231},
          doi = {10.1109/JPROC.2004.840301},
       adsurl = {https://ui.adsabs.harvard.edu/abs/2005IEEEP..93..216F}
}

@ARTICLE{2026MNRAS.tmp.1081H,
       author = {{Horowitz}, Benjamin and {Bayer}, Adrian E.},
        title = "{jFoF: GPU Friends-of-Friends Halo Finding with Gradient Propagation}",
      journal = {\mnras},
         year = 2026,
        month = jun,
          doi = {10.1093/mnras/stag1151},
       adsurl = {https://ui.adsabs.harvard.edu/abs/2026MNRAS.tmp.1081H}
}

@ARTICLE{2025MNRAS.544..960H,
       author = {{Horowitz}, Benjamin and {Luki{\'c}}, Zarija},
        title = "{Differentiable cosmological hydrodynamics for field level inference and high dimensional parameter constraints}",
      journal = {\mnras},
         year = 2025,
        month = nov,
       volume = {544},
       number = {1},
        pages = {960-974},
          doi = {10.1093/mnras/staf1785},
archivePrefix = {arXiv},
       eprint = {2502.02294},
 primaryClass = {astro-ph.CO},
       adsurl = {https://ui.adsabs.harvard.edu/abs/2025MNRAS.544..960H}
}

@ARTICLE{2026A&A...709A.110R,
       author = {{Rossell{\'o}}, P. and {Kitaura}, F.-S. and {Forero S{\'a}nchez}, D. and {Sinigaglia}, F. and {Favole}, G.},
        title = "{Differentiable fuzzy cosmic web for field-level inference}",
      journal = {\aap},
         year = 2026,
        month = may,
       volume = {709},
          eid = {A110},
        pages = {A110},
          doi = {10.1051/0004-6361/202555817},
archivePrefix = {arXiv},
       eprint = {2506.03969},
 primaryClass = {astro-ph.CO},
       adsurl = {https://ui.adsabs.harvard.edu/abs/2026A&A...709A.110R}
}

@ARTICLE{2023arXiv230709504B,
       author = {{Bayer}, Adrian E. and {Seljak}, Uros and {Modi}, Chirag},
        title = "{Field-Level Inference with Microcanonical Langevin Monte Carlo}",
      journal = {arXiv e-prints},
         year = 2023,
        month = jul,
          eid = {arXiv:2307.09504},
        pages = {arXiv:2307.09504},
          doi = {10.48550/arXiv.2307.09504},
archivePrefix = {arXiv},
       eprint = {2307.09504},
 primaryClass = {astro-ph.CO},
       adsurl = {https://ui.adsabs.harvard.edu/abs/2023arXiv230709504B}
}

@ARTICLE{2025JCAP...12..039S,
       author = {{Simon}, Hugo and {Lanusse}, Fran{\c{c}}ois and {de Mattia}, Arnaud},
        title = "{Benchmarking field-level cosmological inference from galaxy redshift surveys}",
      journal = {\jcap},
         year = 2025,
        month = dec,
       volume = {2025},
       number = {12},
          eid = {039},
        pages = {039},
          doi = {10.1088/1475-7516/2025/12/039},
archivePrefix = {arXiv},
       eprint = {2504.20130},
 primaryClass = {astro-ph.CO},
       adsurl = {https://ui.adsabs.harvard.edu/abs/2025JCAP...12..039S}
}

@ARTICLE{2026arXiv260426823D,
       author = {{Dong}, Chenze and {Horowitz}, Benjamin and {Bayer}, Adrian E. and {Lee}, Khee-Gan},
        title = "{Mujic{\ensuremath{\Lambda}}: Reconstructing Initial Conditions from Incomplete Redshift Surveys with Projected Optimization}",
      journal = {arXiv e-prints},
         year = 2026,
        month = apr,
          eid = {arXiv:2604.26823},
        pages = {arXiv:2604.26823},
          doi = {10.48550/arXiv.2604.26823},
archivePrefix = {arXiv},
       eprint = {2604.26823},
 primaryClass = {astro-ph.CO},
       adsurl = {https://ui.adsabs.harvard.edu/abs/2026arXiv260426823D}
}

@ARTICLE{2025MNRAS.540..716M,
       author = {{McAlpine}, Stuart and {Jasche}, Jens and {Ata}, Metin and {Lavaux}, Guilhem and {Stiskalek}, Richard and {Frenk}, Carlos S. and {Jenkins}, Adrian},
        title = "{The Manticore Project I: a digital twin of our cosmic neighbourhood from Bayesian field-level analysis}",
      journal = {\mnras},
         year = 2025,
        month = jun,
       volume = {540},
       number = {1},
        pages = {716-745},
          doi = {10.1093/mnras/staf767},
archivePrefix = {arXiv},
       eprint = {2505.10682},
 primaryClass = {astro-ph.CO},
       adsurl = {https://ui.adsabs.harvard.edu/abs/2025MNRAS.540..716M}
}

@article{ibeid2020fft,
  title={FFT, FMM, and multigrid on the road to exascale: Performance challenges and opportunities},
  author={Ibeid, Huda and Olson, Luke and Gropp, William},
  journal={Journal of Parallel and Distributed Computing},
  volume={136},
  pages={63--74},
  year={2020},
  publisher={Elsevier}
}

@ARTICLE{2025arXiv251213403H,
       author = {{Horowitz}, Benjamin and {Luki{\'c}}, Zarija and {Nagamine}, Kentaro and {Oku}, Yuri},
        title = "{diffhydro: Inverse Multiphysics Modeling and Embedded Machine Learning in Astrophysical Flows}",
      journal = {arXiv e-prints},
         year = 2025,
        month = dec,
          eid = {arXiv:2512.13403},
        pages = {arXiv:2512.13403},
          doi = {10.48550/arXiv.2512.13403},
archivePrefix = {arXiv},
       eprint = {2512.13403},
 primaryClass = {astro-ph.IM},
       adsurl = {https://ui.adsabs.harvard.edu/abs/2025arXiv251213403H}
}

@ARTICLE{2002A&A...385..337T,
       author = {{Teyssier}, R.},
        title = "{Cosmological hydrodynamics with adaptive mesh refinement. A new high resolution code called RAMSES}",
      journal = {\aap},
         year = 2002,
        month = apr,
       volume = {385},
        pages = {337-364},
          doi = {10.1051/0004-6361:20011817},
archivePrefix = {arXiv},
       eprint = {astro-ph/0111367},
 primaryClass = {astro-ph},
       adsurl = {https://ui.adsabs.harvard.edu/abs/2002A&A...385..337T}
}

@article{kabalan2026jaxdecomp,
  title={jaxDecomp: JAX Library for 3D Domain Decomposition and Parallel FFTs},
  author={Kabalan, Wassim and Lanusse, Fran{\c{c}}ois and Boucaud, Alexandre and Aubourg, Eric},
  journal={Journal of Open Source Software},
  volume={11},
  number={121},
  pages={8852},
  year={2026}
}

@inproceedings{romero2022distributed,
  title={Distributed-memory simulations of turbulent flows on modern GPU systems using an adaptive pencil decomposition library},
  author={Romero, Joshua and Costa, Pedro and Fatica, Massimiliano},
  booktitle={Proceedings of the Platform for Advanced Scientific Computing Conference},
  pages={1--11},
  year={2022}
}

@ARTICLE{2024ApJS..270...36L,
       author = {{Li}, Yin and {Modi}, Chirag and {Jamieson}, Drew and {Zhang}, Yucheng and {Lu}, Libin and {Feng}, Yu and {Lanusse}, Fran{\c{c}}ois and {Greengard}, Leslie},
        title = "{Differentiable Cosmological Simulation with the Adjoint Method}",
      journal = {\apjs},
         year = 2024,
        month = feb,
       volume = {270},
       number = {2},
          eid = {36},
        pages = {36},
          doi = {10.3847/1538-4365/ad0ce7},
archivePrefix = {arXiv},
       eprint = {2211.09815},
 primaryClass = {astro-ph.IM},
       adsurl = {https://ui.adsabs.harvard.edu/abs/2024ApJS..270...36L}
}

@article{lin2018python,
  title={Python non-uniform fast Fourier transform (PyNUFFT): An accelerated non-Cartesian MRI package on a heterogeneous platform (CPU/GPU)},
  author={Lin, Jyh-Miin},
  journal={Journal of Imaging},
  volume={4},
  number={3},
  pages={51},
  year={2018},
  publisher={MDPI}
}

@book{poisson1826memoire,
  title={M{\'e}moire sur la th{\'e}orie du magn{\'e}tisme en movement},
  author={Poisson, Sim{\'e}on-Denis},
  year={1826},
  publisher={L'Acad{\'e}mie}
}

@ARTICLE{2017MNRAS.469.1968P,
       author = {{Pan}, Qiaoyin and {Pen}, Ue-Li and {Inman}, Derek and {Yu}, Hao-Ran},
        title = "{Increasing Fisher information by Potential Isobaric Reconstruction}",
      journal = {\mnras},
         year = 2017,
        month = aug,
       volume = {469},
       number = {2},
        pages = {1968-1973},
          doi = {10.1093/mnras/stx774},
archivePrefix = {arXiv},
       eprint = {1611.10013},
 primaryClass = {astro-ph.CO},
       adsurl = {https://ui.adsabs.harvard.edu/abs/2017MNRAS.469.1968P}
}

@ARTICLE{2017ApJ...841L..29W,
       author = {{Wang}, Xin and {Yu}, Hao-Ran and {Zhu}, Hong-Ming and {Yu}, Yu and {Pan}, Qiaoyin and {Pen}, Ue-Li},
        title = "{Isobaric Reconstruction of the Baryonic Acoustic Oscillation}",
      journal = {\apjl},
         year = 2017,
        month = jun,
       volume = {841},
       number = {2},
          eid = {L29},
        pages = {L29},
          doi = {10.3847/2041-8213/aa738c},
archivePrefix = {arXiv},
       eprint = {1703.09742},
 primaryClass = {astro-ph.CO},
       adsurl = {https://ui.adsabs.harvard.edu/abs/2017ApJ...841L..29W}
}

@ARTICLE{1998ApJS..115...19P,
       author = {{Pen}, Ue-Li},
        title = "{A High-Resolution Adaptive Moving Mesh Hydrodynamic Algorithm}",
      journal = {\apjs},
         year = 1998,
        month = mar,
       volume = {115},
       number = {1},
        pages = {19-34},
          doi = {10.1086/313074},
archivePrefix = {arXiv},
       eprint = {astro-ph/9704258},
 primaryClass = {astro-ph},
       adsurl = {https://ui.adsabs.harvard.edu/abs/1998ApJS..115...19P}
}

@ARTICLE{2026arXiv260503563K,
       author = {{Krasnopolsky}, Ruben and {Puro}, Touko and {Li}, Wei-Wen and {Shang}, Hsien and {V{\"a}is{\"a}l{\"a}}, Miikka S. and {Mac Low}, Mordecai-Mark and {Rheinhardt}, Matthias and {Korpi-Lagg}, Maarit},
        title = "{Iterative Poisson Solvers for Self-gravity with the GPU Code Astaroth}",
      journal = {arXiv e-prints},
         year = 2026,
        month = may,
          eid = {arXiv:2605.03563},
        pages = {arXiv:2605.03563},
          doi = {10.48550/arXiv.2605.03563},
archivePrefix = {arXiv},
       eprint = {2605.03563},
 primaryClass = {astro-ph.IM},
       adsurl = {https://ui.adsabs.harvard.edu/abs/2026arXiv260503563K}
}

@INPROCEEDINGS{2018JPhCS1031a2021R,
       author = {{Ramsey}, J.~P. and {Haugb{\o}lle}, T. and {Nordlund}, {\r{A}}.},
        title = "{A simple and efficient solver for self-gravity in the DISPATCH astrophysical simulation framework}",
    booktitle = {Journal of Physics Conference Series},
         year = 2018,
       series = {Journal of Physics Conference Series},
       volume = {1031},
        month = may,
    publisher = {IOP},
          eid = {012021},
        pages = {012021},
          doi = {10.1088/1742-6596/1031/1/012021},
archivePrefix = {arXiv},
       eprint = {1806.10098},
 primaryClass = {astro-ph.IM},
       adsurl = {https://ui.adsabs.harvard.edu/abs/2018JPhCS1031a2021R}
}

@ARTICLE{2026arXiv260510678F,
       author = {{Fischill}, Paul and {Adelmann}, Andreas and {Muralikrishnan}, Sriramkrishnan},
        title = "{A Performance-Portable, Massively Parallel Distributed Nonuniform FFT}",
      journal = {arXiv e-prints},
         year = 2026,
        month = may,
          eid = {arXiv:2605.10678},
        pages = {arXiv:2605.10678},
          doi = {10.48550/arXiv.2605.10678},
archivePrefix = {arXiv},
       eprint = {2605.10678},
 primaryClass = {cs.CE},
       adsurl = {https://ui.adsabs.harvard.edu/abs/2026arXiv260510678F}
}

@ARTICLE{2017JCAP...12..009S,
       author = {{Seljak}, Uro{\v{s}} and {Aslanyan}, Grigor and {Feng}, Yu and {Modi}, Chirag},
        title = "{Towards optimal extraction of cosmological information from nonlinear data}",
      journal = {\jcap},
         year = 2017,
        month = dec,
       volume = {2017},
       number = {12},
          eid = {009},
        pages = {009},
          doi = {10.1088/1475-7516/2017/12/009},
archivePrefix = {arXiv},
       eprint = {1706.06645},
 primaryClass = {astro-ph.CO},
       adsurl = {https://ui.adsabs.harvard.edu/abs/2017JCAP...12..009S}
}

@ARTICLE{2021A&C....3700505M,
       author = {{Modi}, C. and {Lanusse}, F. and {Seljak}, U.},
        title = "{FlowPM: Distributed TensorFlow implementation of the FastPM cosmological N-body solver}",
      journal = {Astronomy and Computing},
         year = 2021,
        month = oct,
       volume = {37},
          eid = {100505},
        pages = {100505},
          doi = {10.1016/j.ascom.2021.100505},
archivePrefix = {arXiv},
       eprint = {2010.11847},
 primaryClass = {astro-ph.CO},
       adsurl = {https://ui.adsabs.harvard.edu/abs/2021A&C....3700505M}
}

@ARTICLE{2013JCAP...06..036T,
       author = {{Tassev}, Svetlin and {Zaldarriaga}, Matias and {Eisenstein}, Daniel J.},
        title = "{Solving large scale structure in ten easy steps with COLA}",
      journal = {\jcap},
         year = 2013,
        month = jun,
       volume = {2013},
       number = {6},
          eid = {036},
        pages = {036},
          doi = {10.1088/1475-7516/2013/06/036},
archivePrefix = {arXiv},
       eprint = {1301.0322},
 primaryClass = {astro-ph.CO},
       adsurl = {https://ui.adsabs.harvard.edu/abs/2013JCAP...06..036T}
}

@ARTICLE{2022ApJS..263...27H,
       author = {{Horowitz}, Benjamin and {Lee}, Khee-Gan and {Ata}, Metin and {M{\"u}ller}, Thomas and {Krolewski}, Alex and {Prochaska}, J. Xavier and {Hennawi}, Joseph F. and {White}, Martin and {Schlegel}, David and {Rich}, R. Michael and {Nugent}, Peter E. and {Suzuki}, Nao and {Kashino}, Daichi and {Koekemoer}, Anton M. and {Lemaux}, Brian C.},
        title = "{Second Data Release of the COSMOS Ly{\ensuremath{\alpha}} Mapping and Tomography Observations: The First 3D Maps of the Detailed Cosmic Web at 2.05 < z < 2.55}",
      journal = {\apjs},
         year = 2022,
        month = dec,
       volume = {263},
       number = {2},
          eid = {27},
        pages = {27},
          doi = {10.3847/1538-4365/ac982d},
archivePrefix = {arXiv},
       eprint = {2109.09660},
 primaryClass = {astro-ph.CO},
       adsurl = {https://ui.adsabs.harvard.edu/abs/2022ApJS..263...27H}
}

@ARTICLE{2020AREPO,
       author = {{Weinberger}, Rainer and {Springel}, Volker and {Pakmor}, R{\"u}diger},
        title = "{The AREPO Public Code Release}",
      journal = {\apjs},
         year = 2020,
        month = jun,
       volume = {248},
       number = {2},
          eid = {32},
        pages = {32},
          doi = {10.3847/1538-4365/ab908c},
archivePrefix = {arXiv},
       eprint = {1909.04667},
 primaryClass = {astro-ph.IM},
       adsurl = {https://ui.adsabs.harvard.edu/abs/2020ApJS..248...32W}
}

@ARTICLE{2024MNRAS.535.1258D,
       author = {{Doeser}, Ludvig and {Jamieson}, Drew and {Stopyra}, Stephen and {Lavaux}, Guilhem and {Leclercq}, Florent and {Jasche}, Jens},
        title = "{Bayesian inference of initial conditions from non-linear cosmic structures using field-level emulators}",
      journal = {\mnras},
         year = 2024,
        month = dec,
       volume = {535},
       number = {2},
        pages = {1258-1277},
          doi = {10.1093/mnras/stae2429},
archivePrefix = {arXiv},
       eprint = {2312.09271},
 primaryClass = {astro-ph.CO},
       adsurl = {https://ui.adsabs.harvard.edu/abs/2024MNRAS.535.1258D}
}

@ARTICLE{2020arXiv201000619V,
       author = {{Villaescusa-Navarro}, Francisco and {Angl{\'e}s-Alc{\'a}zar}, Daniel and {Genel}, Shy and {Spergel}, David N. and {Somerville}, Rachel S. and {Dave}, Romeel and {Pillepich}, Annalisa and {Hernquist}, Lars and {Nelson}, Dylan and {Torrey}, Paul and {Narayanan}, Desika and {Li}, Yin and {Philcox}, Oliver and {La Torre}, Valentina and {Delgado}, Ana Maria and {Ho}, Shirley and {Hassan}, Sultan and {Burkhart}, Blakesley and {Wadekar}, Digvijay and {Battaglia}, Nicholas and {Contardo}, Gabriella},
        title = "{The CAMELS project: Cosmology and Astrophysics with MachinE Learning Simulations}",
      journal = {arXiv e-prints},
         year = 2020,
        month = oct,
          eid = {arXiv:2010.00619},
        pages = {arXiv:2010.00619},
archivePrefix = {arXiv},
       eprint = {2010.00619},
 primaryClass = {astro-ph.CO},
       adsurl = {https://ui.adsabs.harvard.edu/abs/2020arXiv201000619V}
}

@ARTICLE{2019TARDIS,
    author = {{Horowitz}, Benjamin and {Lee}, Khee-Gan and {White}, Martin and {Krolewski}, Alex and {Ata}, Metin},
        title = "{TARDIS. I. A Constrained Reconstruction Approach to Modeling the z {\ensuremath{\sim}} 2.5 Cosmic Web Probed by Ly{\ensuremath{\alpha}} Forest Tomography}",
      journal = {\apj},
         year = 2019,
        month = dec,
       volume = {887},
       number = {1},
          eid = {61},
        pages = {61},
          doi = {10.3847/1538-4357/ab4d4c},
archivePrefix = {arXiv},
       eprint = {1903.09049},
 primaryClass = {astro-ph.CO},
       adsurl = {https://ui.adsabs.harvard.edu/abs/2019ApJ...887...61H}
}

@article{2018BORG,
	author = {{Jasche}, J. and {Lavaux}, G.},
        title = "{Physical Bayesian modelling of the non-linear matter distribution: New insights into the nearby universe}",
      journal = {\aap},
         year = 2019,
        month = may,
       volume = {625},
          eid = {A64},
        pages = {A64},
          doi = {10.1051/0004-6361/201833710},
archivePrefix = {arXiv},
       eprint = {1806.11117},
 primaryClass = {astro-ph.CO},
       adsurl = {https://ui.adsabs.harvard.edu/abs/2019A&A...625A..64J}
}

@article{fastPM,
	Adsurl = {http://adsabs.harvard.edu/abs/2016MNRAS.463.2273F},
	Archiveprefix = {arXiv},
	Author = {{Feng}, Y. and {Chu}, M.-Y. and {Seljak}, U. and {McDonald}, P.},
	Doi = {10.1093/mnras/stw2123},
	Eprint = {1603.00476},
	Journal = {\mnras},
	Month = dec,
	Pages = {2273-2286},
	Title = {{FASTPM: a new scheme for fast simulations of dark matter and haloes}},
	Volume = 463,
	Year = 2016}




\appendix

\section{Error accumulation of the warm-start solve}
\label{app:accum}

A finite-cycle iterative solver leaves a small residual at every step, and because that residual
perturbs the particle positions it feeds forward into the source of the next step. A reasonable concern
is whether this per-step solver error accumulates coherently along the trajectory and eventually
biases the evolved field. Here we isolate that effect directly and show that in the cases studied in this work it does not. The
warm-start multigrid error saturates rather than growing, so a small fixed cycle budget is sufficient
for the whole run.

\subsection{Experiment.} We evolve two particle--mesh trajectories from identical initial
conditions. The reference trajectory solves the
discrete Poisson equation \eqref{eq:linsys} exactly at every step with the spectral method; the test
trajectory uses the warm-start multigrid solver with a fixed budget of a single V-cycle (Chebyshev
smoother, $\alpha=8$), warm-started from the growth-scaled previous potential \eqref{eq:growth}. Both
trajectories use the same discrete operator and the same real-space fourth-order force gradient, so
the difference between them is due purely to the finite-cycle convergence of the multigrid solve and
not to any difference in discretization or force computation. The run is a $256^3$ mesh in a
$256\,h^{-1}\mathrm{Mpc}$ box, integrated with a fixed-step leapfrog over $30$ steps from $a=0.1$ to
$a=1$. The multigrid trajectory is seeded once at the first step with a deep solve, whose potential
agrees with the spectral solve to a relative $L_2$ of $1.4\times10^{-4}$.

At each step we quantify the error in two ways. The injected residual is the instantaneous
relative residual $\norm{F-A\vphi}/\norm{F}$ of the multigrid solve at that step, i.e. the error added to
the potential before it propagates. The accumulated error is the difference between the two
trajectories' potentials, $\vphi_{\rm MG}-\vphi_{\rm FFT}$, where for a clean comparison the potential
of both evolved particle distributions is re-evaluated with the same exact solver. This
difference therefore reflects genuine divergence of the two trajectories rather than the
representation of the potential.

\begin{figure*}
    \centering
    \includegraphics[width=0.99\linewidth]{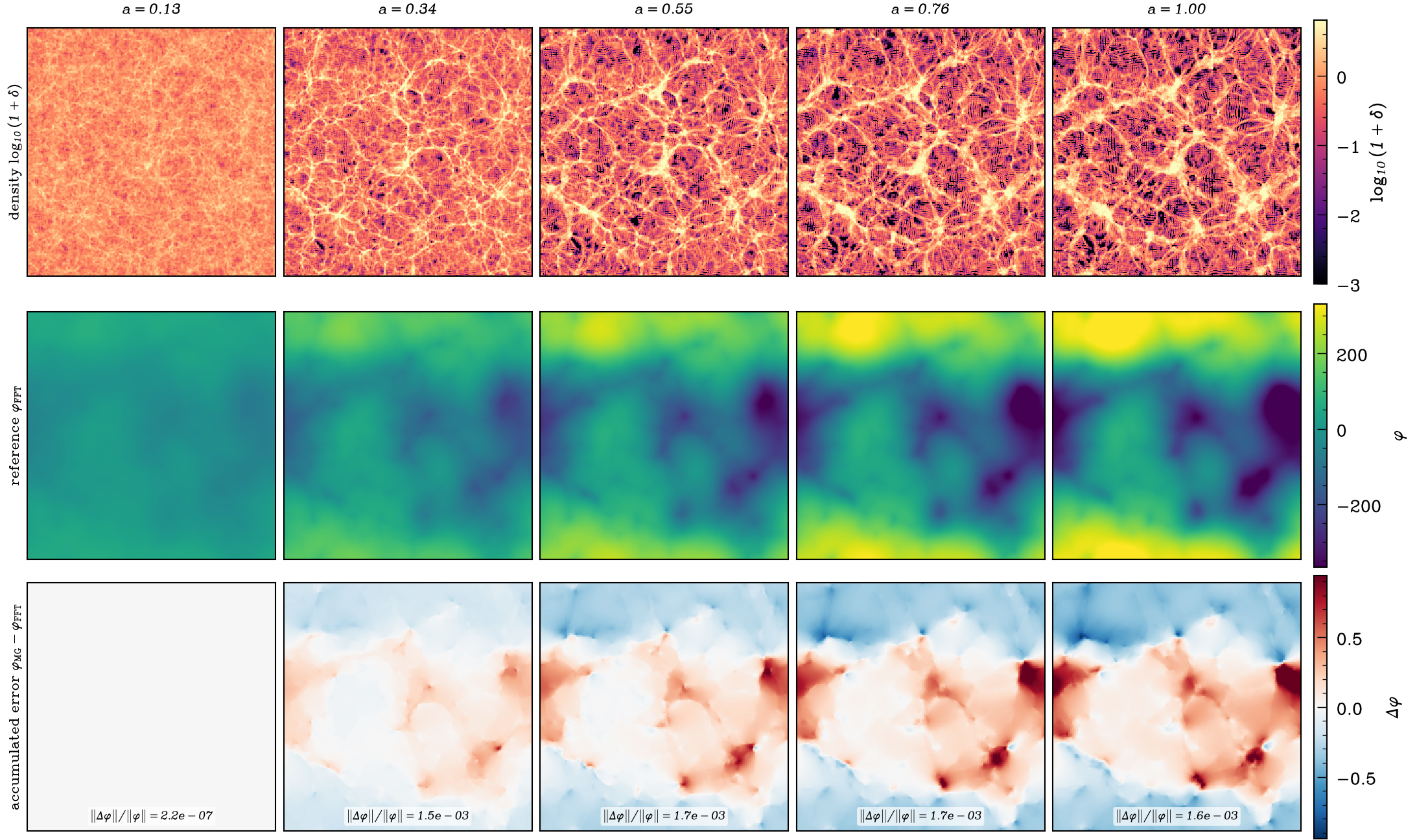}
    \caption{Accumulated error of the warm-start multigrid gravity solve over a $256^3$ particle--mesh
    run, at scale factors $a=0.13,0.34,0.55,0.76,1.00$. \emph{Top:} the evolving density field
    $\log_{10}(1+\delta)$, showing the cosmic web forming from near-uniform initial conditions.
    \emph{Middle:} the reference (spectral) potential $\vphi_{\rm FFT}$ of the same distribution.
    \emph{Bottom:} the accumulated potential error $\vphi_{\rm MG}-\vphi_{\rm FFT}$ between two
    trajectories evolved from identical initial conditions that differ only in the gravity solve
    (exact spectral versus a single warm multigrid V-cycle with the Chebyshev smoother), on a fixed
    colour scale with the relative $L_2$ annotated per panel. The error is small ($\sim\!10^{-3}$) and
    tracks the collapsed structure in the density field, being largest on the deepest potential wells.}
    \label{fig:accum_slices}
\end{figure*}

\begin{figure}
    \centering
    \includegraphics[width=0.99\linewidth]{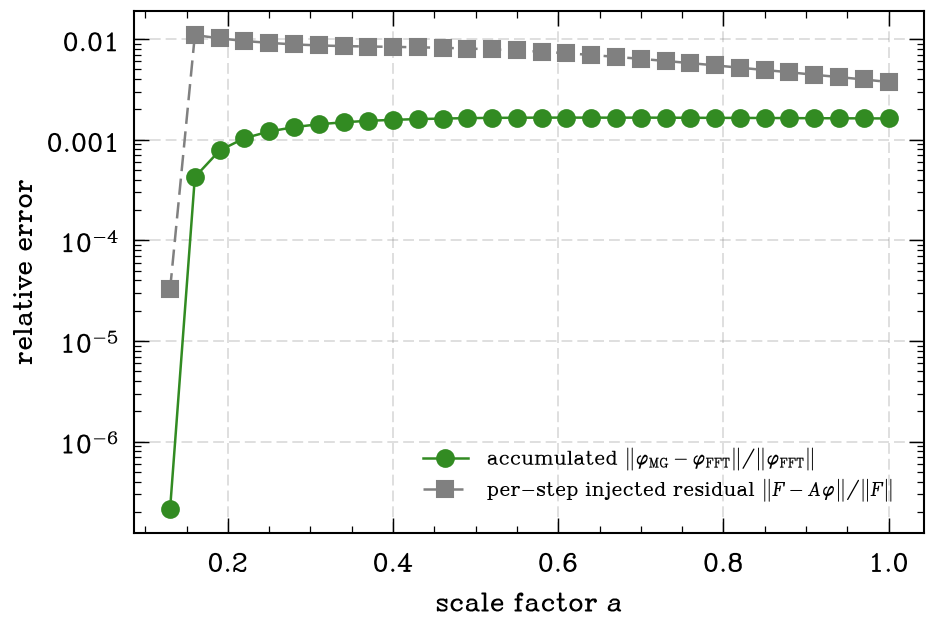}
    \caption{Growth of the two error measures from Fig.~\ref{fig:accum_slices} as a function of scale
    factor. The accumulated field error $\norm{\vphi_{\rm MG}-\vphi_{\rm FFT}}/\norm{\vphi_{\rm FFT}}$
    (red) rises over the first few steps and then saturates near $1.6\times10^{-3}$, while the
    per-step injected residual $\norm{F-A\vphi}/\norm{F}$ (grey) stays five to ten times larger. The
    accumulated error remaining below, and not growing with, the per-step residual demonstrates that
    the warm-start solver errors do not add coherently along the trajectory: the scheme is
    self-correcting, so a fixed small cycle budget suffices for the entire run.}
    \label{fig:accum_growth}
\end{figure}

\subsection{Results.} Figure~\ref{fig:accum_slices} shows midplane slices of the evolving density
(top), the reference potential (middle), and the accumulated potential error (bottom) at five scale
factors. The error grows during the first several steps as the two trajectories begin to diverge, but
it remains small (a relative $L_2$ of order $10^{-3}$) and, crucially, it is spatially localized to
the collapsed structures forming in the density field. The largest excursions sit on the deepest
potential wells (the massive halo at the right of each panel), consistent with the field-level result
of \S\ref{sec:results} that the residual is concentrated in the dense cores of collapsed regions,
where the discrete operator and the finite cycle budget differ most. The cosmic-web structure itself
is reproduced faithfully.

Figure~\ref{fig:accum_growth} makes the accumulation behavior quantitative. The accumulated field
error rises over the first ${\sim}10$ steps to a plateau at $\approx1.6\times10^{-3}$ and then stops
growing---indeed it decreases slightly toward $a=1$---even as the per-step injected residual remains
five to ten times larger throughout ($\sim\!10^{-2}$ early in the run, falling to ${\sim}4\times10^{-3}$
by the present day as the warm-start predictor improves with the shrinking relative inter-step change).
That the accumulated error sits well below the per-step residual, and saturates rather than growing
linearly with the number of steps, is the signature of a self-correcting scheme: the solver errors
injected at successive steps do not add coherently, because each warm solve drives the residual back
below the level set by the integrator's own truncation error. The consequence for production runs is
that the accuracy of the evolved field is controlled by the (small, fixed) per-step cycle budget and
does not degrade with trajectory length; the accuracy/cost trade-off of \S\ref{sec:results} therefore
holds over the full evolution.


\section{Moving-Mesh Grid Limiters}
\label{sec:limiters}

Left unchecked, the inward flow of the moving mesh shown in Eq.~\ref{eq:calcdefp} would eventually drive cells
to zero volume ($\sqrt{g}\to0$), folding the mesh and rendering the geometry
singular. The method therefore stabilises the deformation with a layered set of
limiters, applied at three stages: to the source of the mesh solve, as a
post--solve repulsive correction, and as a final hard rescaling. The limiters are
controlled by two parameters, a maximum admissible compression $c_{\max}$ and a
maximum admissible anisotropy (skew) $s_{\max}$. They are designed to act
locally, intervening only in the few cells that approach the limits rather
than freezing the entire mesh.

\subsection{Eigenvalue bounds}
Several of the limiters and the mesh stability criterion require the eigenvalues of
the symmetric triad on every cell. By default we compute them exactly from
the symmetric $3\times3$ triad, which is affordable as a batched, vectorised
per-cell operation on the GPU and gives the tightest limiter response. Where only a
one--sided guarantee is needed, or to trade accuracy for cost, the spectrum can
instead be bracketed by the Gershgorin disc bounds,
\begin{equation}
  \lambda_{\min}^{\rm lb} = \min_{\alpha}\Big(\Amat_{\alpha\alpha}
      - \!\!\sum_{\beta\neq\alpha}\!|\Amat_{\alpha\beta}|\Big),
  \qquad
  \lambda_{\max}^{\rm ub} = \max_{\alpha}\Big(\Amat_{\alpha\alpha}
      + \!\!\sum_{\beta\neq\alpha}\!|\Amat_{\alpha\beta}|\Big),
  \label{eq:gersh}
\end{equation}
which bound the true spectrum from one side at negligible cost and are themselves
smooth functions of the field.

\subsection{Source--side limiting}
Acting on the right--hand side of \eqref{eq:calcdefp}, the normalised density
deficit $t=\tilde\rho_{\rm init}-\tilde\rho$ (with $\tilde\rho=\rho\sqrt{g}$
rescaled to unit mean) is first clipped in amplitude and relaxed over a
characteristic number of steps, then smoothed,
\begin{equation}
  t \;\longleftarrow\; \frac{\operatorname{clip}(t,-x_{\rm hi},x_{\rm hi})}{n_{\rm relax}\,\dd\tau_{\rm old}},
  \qquad f = \kappa\,\mathrm{smooth}(t).
  \label{eq:cliptmp}
\end{equation}
Two protective switches then act on $f$. Where a cell has become strongly
over--expanded, the flux is reversed so as to drive it back toward the target
density and where a cell is close to its compression limit, only further expansion
is permitted:
\begin{equation}
  f \;\longleftarrow\;
  \begin{cases}
    -2\,|f| & \sqrt{g} > 10\,\tilde\rho_{\rm init} \quad(\text{over--expanded}),\\[2pt]
    \max(f,0) & \lambda_{\min}^{\rm lb} < 1/c_{\max}^{\,\rm loc} \quad(\text{near collapse}),\\[2pt]
    f & \text{otherwise.}
  \end{cases}
  \label{eq:rhslim}
\end{equation}
Finally the source is clipped to $f\in[-1/\dd\tau,\,1/\dd\tau]$, which bounds the
mesh displacement to at most about one cell per step and thereby preserves the
validity of the explicit update. Here the local compression limit is scaled by
the target density as $c_{\max}^{\,\rm loc}=c_{\max}\,\tilde\rho_{\rm init}^{-1/3}$,
and $\lambda_{\min}^{\rm lb}$ is the Gershgorin bound \eqref{eq:gersh} of the
current triad.

\subsection{Repulsive post--solve correction}
After the deformation--rate solve, a further safeguard anticipates collapse on the
candidate mesh $\psi+\dd\tau_{\rm old}\dot\psi$, i.e. on the geometry that
the step is about to produce. Let $\lambda_{\min}^{\rm lb}$ now denote the
Gershgorin bound of the candidate triad and
$\lambda_{\min}^{\rm req}=1/c_{\max}^{\,\rm loc}$ the minimum admissible
eigenvalue. A repulsive potential is generated wherever the candidate would
violate the bound and is added back to the deformation rate,
\begin{equation}
  t_{\rm lim} = \max\!\Big(0,\ \frac{3\,(\lambda_{\min}^{\rm req}-\lambda_{\min}^{\rm lb})}
                                   {\dd\tau_{\rm old}}\Big),
  \qquad
  \nabla^2 c = t_{\rm lim},
  \qquad
  \dot\psi \;\longleftarrow\; \dot\psi + \eta\, c,
  \label{eq:repulsive}
\end{equation}
where $\nabla^2$ is the flat (uniform--grid) Laplacian, solved by FFT, and $\eta$
sets the strength of the correction. Because the correction is sourced only by the
cells that would otherwise over--compress, it perturbs the global flow minimally
while halting incipient folding.

\subsection{Hard rescaling}
As a final guarantee, a scaling factor $s\in[0,1]$ is determined---globally, or
per cell for locality---such that the updated triad remains well conditioned. With
$\Amat(s)=\mathbb{I}+\Hess(\psi+s\,\dd\tau\,\dot\psi)$ and exact eigenvalues
$\lambda_k(s)$, one seeks the largest admissible step,
\begin{equation}
\begin{aligned}
  s^\star=&\max\Big\{s\in[0,1]:\;
  \tfrac{1}{c_{\max}}\le\lambda_{\min}(s),\\
  &\lambda_{\max}(s)\le c_{\max},\quad
  \tfrac{\lambda_{\max}(s)}{\lambda_{\min}(s)}\le s_{\max}\Big\},\\
  &\dot\psi\longleftarrow s^\star\dot\psi .
\end{aligned}
\label{eq:bisection}
\end{equation}
The maximiser is found by a fixed number of bisection steps, and the procedure
fails safe, reverting to no mesh motion if any eigenvalue is non--finite. In
normal operation the source--side and repulsive limiters keep the mesh well
inside these bounds, and the hard rescaling rarely activates. It exists to
guarantee a valid mesh under all circumstances.

\subsection{Mesh time--step constraint}
The explicit mesh update is itself subject to a stability condition; i.e. the geometry
should not change too rapidly within one step. Since the rate of change of the
triad is $\dot{\Amat}\approx\Hess\dot\psi$, an admissible step follows from
bounding the spectral radius of $\Hess\dot\psi$, again via a Gershgorin estimate,
\begin{equation}
  \dd\tau_{\rm mesh} \;<\; \frac{C_{\rm cfl}}{5\,\max_{\xixi}\big|\lambda(\Hess\dot\psi)\big|},
  \label{eq:meshcfl}
\end{equation}
with a safety factor $C_{\rm cfl}<1$. The global time--step is the smaller of this
mesh constraint and the usual constraint from the particle dynamics. In order to make direct comparison against the static mesh approach, we do not explicitly enforce this criteria in the examples in the main paper and instead use a fixed timestepping schedule.


\bsp	
\label{lastpage}
\end{document}